\documentclass{aa}  

\usepackage{graphicx}
\usepackage{subcaption}
\usepackage{siunitx}
\usepackage{booktabs}
\usepackage{placeins}  
\usepackage{hyperref} 
\usepackage{multirow}
\usepackage{booktabs}
\usepackage{array}
\usepackage{sidecap}

\begin{document}

   \title{Spectroscopic analysis of hydrogen and silicon in bright fireballs: New insights into meteoroid composition}
   \titlerunning{Spectroscopic analysis of H and Si in fireballs}
   \subtitle{}

\author{V. Voj\'{a}\v{c}ek
          \inst{1}
          \and
          J. Borovi\v{c}ka\inst{1}
         \and
         P. Spurn\'{y}\inst{1}
          }
\authorrunning{Voj\'{a}\v{c}ek et al.}
\institute{Astronomical Institute of the Czech Academy of Sciences,
               Fri\v{c}ova 298, 251 65 Ond\v{r}ejov, Czech Republic\\
              \email{vojacek@asu.cas.cz}
             }

   \date{}


  \abstract
   {}
   {We present a study of the high-temperature spectral component in meteor fireballs, with a particular focus on neutral hydrogen (H$_\alpha$ at 656.28 nm) and ionised silicon (Si~II~-~2 doublet at 634.71 nm and 637.14 nm). By analysing  spectra from the European Fireball Network (EN) that exhibit H$_\alpha$ and Si~II~-~2 emissions, we investigated the relationship between hydrogen and silicon abundances across different meteoroid types. The plasma temperature of the high-temperature component  remains independent of meteor velocity. This allows us to directly compare relative intensities of hydrogen, bound in more volatile materials, with silicon, bound in less volatile materials, in bodies with different velocities.}
   {We analysed 31 meteor spectra from the EN, focusing on H$_\alpha$ (656.28 nm) and Si~II~-~2 (634.71 nm, 637.14 nm) emissions to determine the elemental abundances and their relationships with the meteor parameters. The spectroscopic data were reduced following established procedures to derive the line intensities. We employed direct line integration and applied ionisation corrections through Saha equations to estimate the relative atomic abundances.}
   {Our results confirmed that the H/Si value remains largely independent of meteor velocity. We show a positive correlation with photometric mass for cometary meteoroids, suggesting that larger bodies better preserve their volatile content, namely hydrogen.  This correlation persists across the meteor showers, showing a physical process related to volatile preservation rather than specific parent body composition. Our data suggest that the abundance of hydrogen in large cometary meteoroids is not only higher than in CI chondrites, but is also comparable to or higher than the measured abundances in small particles of dust from  Halley's comet, depending on the assumed plasma conditions. This work brought new constraints on the distribution and preservation of volatile elements in Solar System bodies and new insights into the potential delivery mechanisms of water to Earth.}
   {The H/Si values show no correlation with meteor velocity, but increase with photometric mass for cometary meteoroids. The prevalence of hydrogen in larger cometary meteoroids supports models where comets could be significant contributors to Earth's volatile inventory.}


   \maketitle

\section{Introduction}\label{introduction}

Meteor spectroscopy is a valuable tool for studying the chemical composition of meteoroids that completely ablate in the atmosphere. When combined with known orbital parameters, chemical data can reveal crucial information about the composition, distribution, and evolution of the smallest bodies in the Solar System and their parent bodies: comets and asteroids. A mix of meteoritic material and atmosphere gases in the form of radiating plasma is created around the flying body. Typically, meteors are observed in visible, near-IR, and near-UV spectral ranges. In this radiation, individual atomic lines and molecular bands can be identified.

\begin{figure}
   \centering
   \includegraphics[width=\hsize]{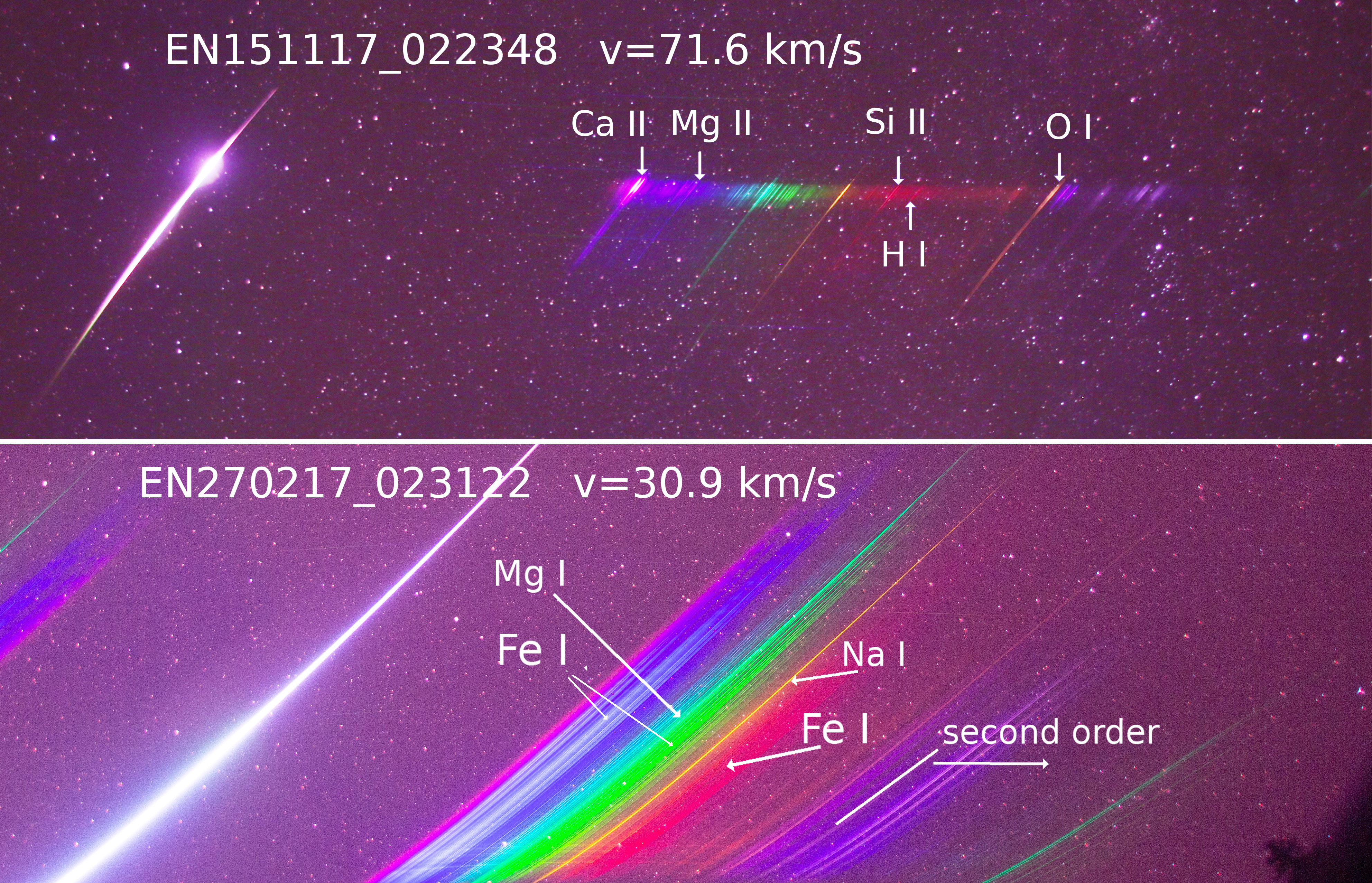}
      \caption{Two spectra with different meteor velocities. The bottom panel shows a relatively slow meteor with the low-temperature component in the spectrum. The top panel shows a fast
meteor with the high-temperature components marked.}
         \label{differenceExample}
   \end{figure}

Meteor spectra exhibit two distinct components that correspond to different temperature regimes \citep{Borovicka1994b}. The first is a low-temperature component (4000-5000~K) originating in the meteor head, characterised by emissions mainly from neutral atoms of sodium, magnesium, iron, manganese, chromium, and singly ionised calcium. The second is a high-temperature component (approximately 10000~K) believed to originate in the shock wave in front of the meteor. In this component we can observe emissions from neutral hydrogen and oxygen atoms, singly ionised silicon, and magnesium or calcium. The differences between a fast meteor with a high-temperature component and a slower meteor with only a low-temperature component in its spectrum are shown in Figure \ref{differenceExample}. The velocity data for these two fireballs were taken from \cite{Borovicka2022a}.

The high-temperature component becomes observable only in meteors with velocities exceeding $30$~km/s, where sufficient kinetic energy exists to heat enough plasma to high temperature. This component typically constitutes a minor portion of the overall meteor spectrum. The relative intensities of low-temperature component lines exhibit a significant dependence on meteor velocity \citep{borovicka2005}. This velocity-induced variation originates from corresponding temperature changes. The meteor head temperature can range from approximately $4000$~K for slow meteors ($15$~km/s) to $5000$~K for high-speed meteoroids. Consequently, this temperature dependence selectively favours certain emission lines. For slow meteors, low-excitation lines (e.g. sodium) predominate. As the meteoroid velocity increases, the higher kinetic energy enables atoms with greater excitation potential to radiate. In contrast, the high-temperature component maintains a consistent temperature of approximately $10000$~K across all meteor velocities \citep{Borovicka1994b}. This temperature invariance means that the relative intensities of high-temperature component lines are independent of velocity, thereby allowing the direct analysis of this spectral component without velocity-induced complications.

One interesting spectral feature is the hydrogen line at $656.28$ nm, commonly known as the H$_\alpha$ line, first confirmed in meteor spectra by \citet{Millman1953} in Perseid spectra.
Other studies provided initial insights into the presence of the H$_\alpha$ line in meteor spectra, demonstrating its occurrence in meteoroids. \citet{BorovickaBetlem1997} studied two bright Perseid spectra and measured a hydrogen abundance comparable to or lower than in carbonaceous chondrites, and significantly lower than in comet Halley dust. \citet{BorovickaJenniskens2000} observed a higher abundance of hydrogen in the Leonids,  compared to a previous study of the Perseids. \citet{Jenniskens2004Hydrogen} measured hydrogen lines in bright meteor spectra and concluded that the H$_\alpha$ line has origin in meteor material and not in atmospheric water molecules. \citet{Matlovic2022} conducted a large-scale survey, presenting comprehensive data on hydrogen emission from both meteors and ablated meteorites. They observed hydrogen lines in meteors faster than $30$~km/s, and confirmed that this line has origin in the high-temperature component.  They found that 92\% of the $62$ studied meteors with hydrogen lines had cometary orbits. They also observed hydrogen depletion for meteors with perihelia smaller than $0.4$~AU. Additionally, they studied real meteorites with simulated ablation in a wind tunnel and analysed the meteor spectra with the high-resolution Echelle spectrograph, observing the strongest hydrogen emission from carbonaceous chondrites and some achondrites. They concluded that H$_\alpha$ emission correlates with other volatiles (Na and CN) and can be viewed as a suitable tracer of water and organics in meteoroids. They compared H$_\alpha$ only with Mg, Na, and Fe lines of low-temperature component. This involved the dependence of the relative H$_\alpha$ brightness on meteor velocity and as a consequence complication in interpretation. They also did not compute atomic abundances.

Hydrogen in interplanetary material is now of great interest as the high water abundance on Earth and the origin of organics—both building blocks of life on Earth—are fundamental questions for understanding the origin of life on our planet. Many primitive asteroids once contained abundant water, which is now stored as OH in hydrated minerals \citep{Alexander2012}. Hydrogen abundance has been studied in carbonaceous chondrites \citep{Joy2020}, in enstatite chondrites \citep{Piani2020}, in samples of several meteoritic materials \citep{Patzek2020}, in asteroidal material \citep{Praet2021a, Praet2021b}, and in cometary material \citep{Hoppe2023}. Most of these papers compare the deuterium to hydrogen ratio (D/H) in studied material with the D/H of Earth's water in an effort to find the most likely source of water on Earth.

It is believed that Earth was initially very hot, thus causing all water to evaporate. Water is therefore thought to be a later addition after the planet's formation. One hypothesis suggests that comets or carbonaceous chondrites from the outer Solar System delivered water to Earth. The D/H values measured in water from different Solar System objects show considerable variability \citep{Altwegg2015}. Asteroids generally show D/H values similar to that of Earth's ocean water, while long-period comets from the Oort cloud show higher deuterium abundance. Among Jupiter-family comets, comet 103P/Hartley 2 has a D/H  very similar to that of Earth's water, while another Jupiter-family comet, 67P/Churyumov-Gerasimenko, showed a value that is 4.5 times higher, even higher than most long-period comets \citep{Altwegg2015}. These findings raised questions about   the origin scenarios of different comet families and the origin of Earth's water. However, \citet{Mandt2024} found that the dust released from comet 67P increased the D/H. Outside of the influence of dust, the ratio is only 1.2 to 1.6 times the terrestrial value, reopening the possibility that Jupiter-family comets could have contributed a significant fraction of Earth's water. \citet{Piani2020} measured hydrogen content and deuterium-to-hydrogen ratios in enstatite chondrite meteorites and found far more hydrogen than previously thought, with values close to that of Earth's mantle. Combining their measurements with cosmochemical models, they showed that most of the Earth's water could have formed from hydrogen delivered by enstatite chondrite meteorites. Whether the water was added to Earth from the inner or outer part of the Solar System, meteoroids were crucial bearers of water, hydrogen, oxygen, or carbon, all important building stones for life formation.

For this study we investigated the high-temperature spectral component of meteor fireballs, with particular emphasis on neutral hydrogen at $656.28$~nm (H$_\alpha$) and the ionised silicon doublet at $634.71$~nm and $637.14$~nm. Our research addresses the following specific questions:

1. What the relationship is between hydrogen and silicon abundances in bright fireballs of different origins (cometary, asteroidal, shower members, and sporadic meteors).

2. How the relative intensities of H$_\alpha$ and Si~II~-~2 lines vary with meteor velocity, brightness, and photometric mass.

3. Whether there is a correlation between  H$_\alpha$ and Si~II~-~2 ratios and meteoroid orbital parameters, particularly for bodies from different source regions of the Solar System.

4. Whether meteor showers display characteristic and consistent  H$_\alpha$ and Si~II~-~2 signatures or whether there is significant variation within each shower population.

5. What the implications are of our findings for understanding the volatile content of different Solar System bodies and potentially for the delivery of water and organics to Earth.

By studying hydrogen and silicon in fireball spectra using observations from the European Fireball Network (EN), we can examine these questions for meteors with brightness ranging from $-10$ to $-14$ magnitude, equivalent to meteoroid sizes of centimetres to decimetres.

\section{Instrumentation and data reduction}\label{instrumentation}

\begin{figure}
   \centering
   \includegraphics[width=\hsize]{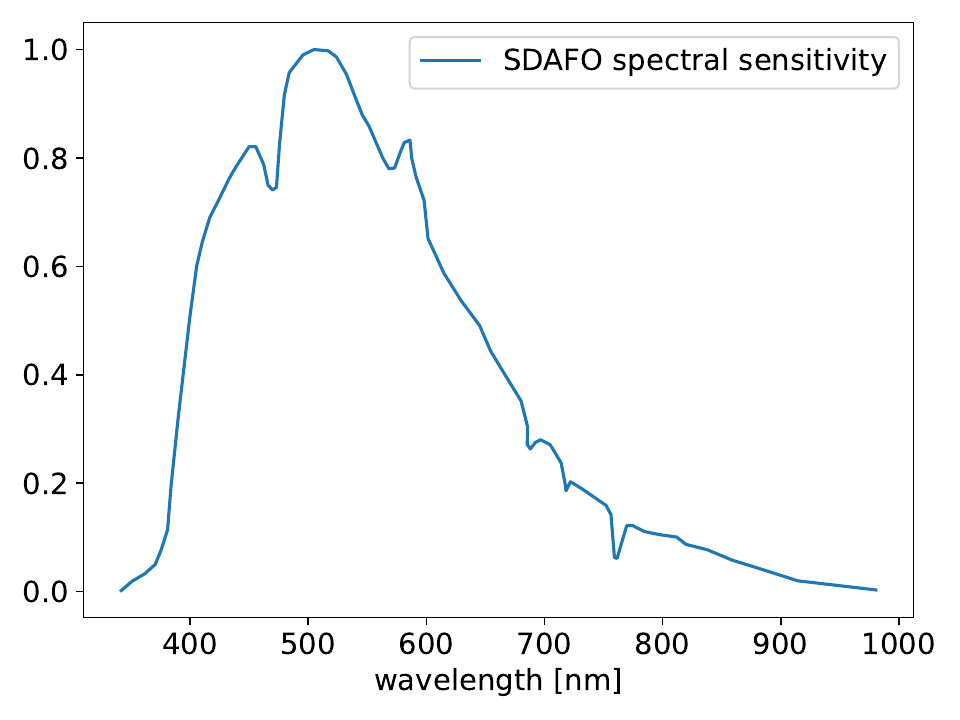}
      \caption{Normalised spectral sensitivity of SDAFO cameras. The digital camera response was combined with atmospheric absorption (telluric absorption lines).}
         \label{FigSensitivity}
   \end{figure}

For this work we used spectral photographic observations of fireballs. The spectral database of observations from Spectral Digital Autonomous Fireball Observatories (SDAFO), which belong to the European Fireball Network \citep{Borovicka2019IMC}, was searched for cases with possible hydrogen or silicon emission in the $630 - 660$~nm region. The SDAFO incorporates two DSLR Canon 6D cameras with their UV/IR filters removed, enhancing the spectral sensitivity range from $380$~nm to $900$~nm (Figure \ref{FigSensitivity}). These cameras are equipped with a Sigma 15mm f/2.8 EX DG Fisheye lens and a plastic holographic grating with 1000 grooves per mm placed in front of the lens. This setup offers a field of view of $100^\circ$ by $77^\circ$ and a spectral resolution of $1.2$~nm per pixel. The system is capable of observing meteors with brightness ranging from -8 to -15 magnitude. Cameras are positioned perpendicular to each other, so if the meteor flight is along the dispersion direction of one camera (and thus spectral lines overlap each other), the second camera captures a resolved spectrum. SDAFO cameras take exposures of $30$ seconds  with 1--2 second pauses between exposures. To cover this pause, exposures in cameras are shifted by $15$ seconds from each other. The lens aperture is set to the native value. Only the ISO is changed from 200 to 800 according to the Moon phase and visibility of the Moon.

The European Fireball Network consists of $20$ stations (as of 2025) equipped with Digital Autonomous Fireball Observatories (DAFO). These cameras are essentially non-spectral versions of SDAFO, equipped with Sigma 8mm F3.5 EX DG Circular Fisheye lenses, an LCD shutter on the back of the lens to measure fireball speed, and a sensitive radiometer for capturing high-resolution light curves \citep{Spurny2017}. Observations from these cameras were used for the meteor orbit and atmospheric trajectory computations. These computations are achieved with our standard procedure described in \cite{Borovicka2022a}, providing high precision of meteor parameters. Moreover, the event time from automatic fireball detection in DAFO images is used for manual identification of their spectral counterparts. The first SDAFO camera was installed in 2015, and as of 2025, they are part of $11$ stations of the European Fireball Network. The number of stations gradually increased during the time frame when data collection for this work began.

In this work, we focused on well-resolved high-temperature component spectral lines, specifically the neutral hydrogen H~I at $656.28$~nm (H$_\alpha$) and the ionised silicon doublet Si~II~-~2 at $634.71$~nm and $637.14$~nm. The upper excitation potential of the hydrogen is $12$~eV, and the upper excitation potential of the ionised silicon lines is $10$~eV. The ionisation potential of hydrogen is $13.6$~eV, and the ionisation potential of silicon is $8.15$~eV. Of the approximately 1100 meteors in the database (as of May 2025), observed by SDAFO between 2015 and 2024, 45 cases exhibiting line radiation between $630$~nm and $660$~nm were manually selected, indicating the potential presence of H~I and Si~II~-~2 lines.

The naming convention reflects the date and time (in UT) of the meteor observation and is as follows:  $ENddmmyy\_hhmmss$ ($EN$ -- European Fireball Network, $dd$ - day,  $mm$ - month, $yy$ --  last two digits of the year,  $hhmmss$ -- time (UT) of
the observed fireball).

Selected meteor spectra were reduced following the process described in \citet{Segon2024}. The dark frame is subtracted, and the image is corrected for camera and lens vignetting by division with the wavelength-independent flat-field. The spectrum is corrected for the spectral grating incidence angle, which affects both absolute and wavelength-dependent brightness changes in the spectrum intensity, using the wavelength-dependent flat-field. Wavelength-dependent atmospheric extinction is also corrected. For short meteors, the whole spectrum is scanned; for long meteors, only part of the spectrum (usually the brightest part) is scanned. Wavelength is calibrated using the wavelengths of known and separated lines with polynomial interpolation. The spectral sensitivity of the system was measured using spectra of the Moon and Venus. The spectral sensitivity curve includes telluric lines, as shown in Figure \ref{FigSensitivity}.

After these reduction processes, we identified the spectral lines of interest. Meteors with no detection of both H$_\alpha$ and Si~II~-~2 were omitted from further investigation.

After line identification, we obtained two groups of meteor spectra. One group in which both Si~II~-~2 and H$_\alpha$ were clearly visible, well above the noise level ($18$ meteors). The second group consisted of cases where Si~II~-~2 was visible, but the H$_\alpha$ was difficult to detect with a signal-to-noise ratio (S/N) close to $1$ (four meteors) or not visible at all (nine meteors). In these cases, we measured the level of the noise, and this quantity marks a possible upper limit for the hydrogen radiation.

   \begin{figure}
   \centering
   \includegraphics[width=\hsize]{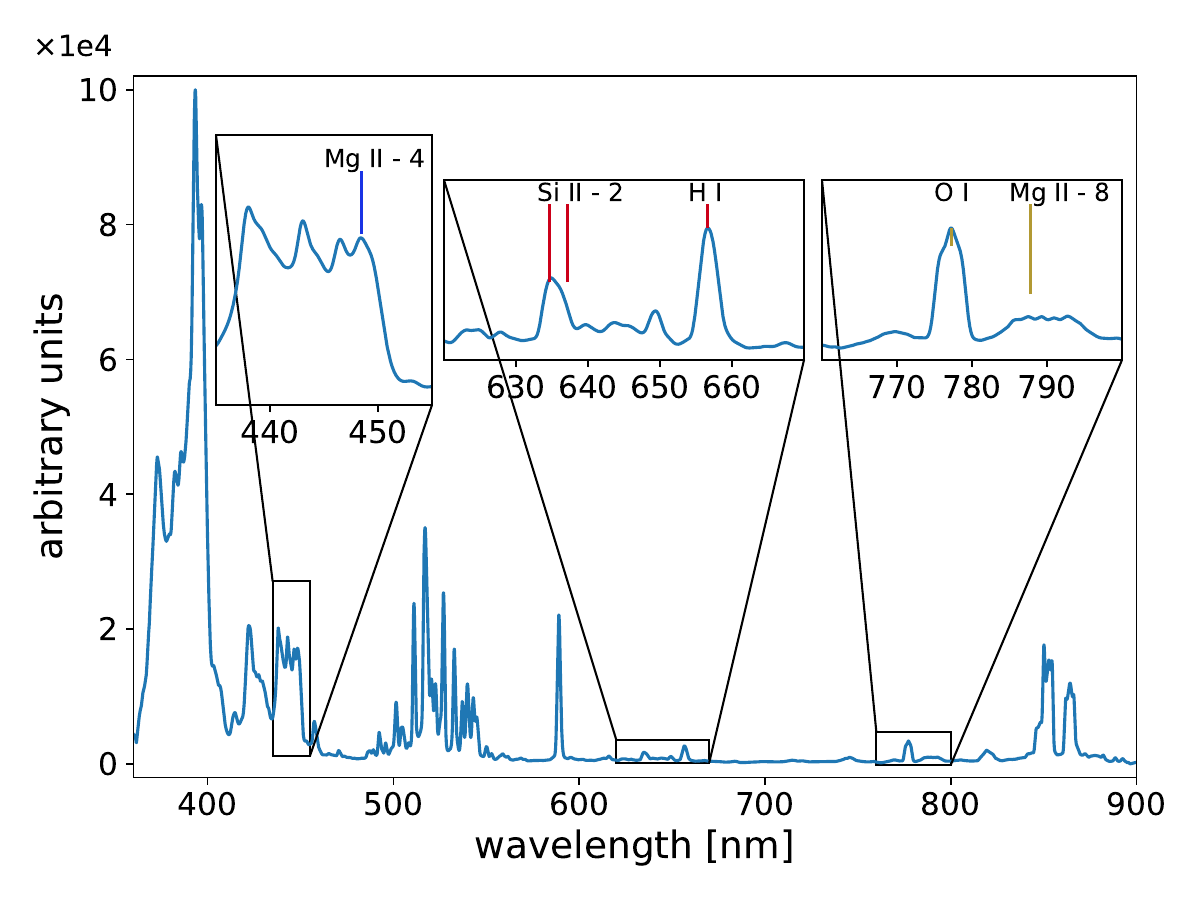}
      \caption{Reduced Lyrid meteor spectrum EN230421\_013038. Zoomed-in images of the analysed spectral regions are shown in the insets, and the positions of the studied high-temperature component lines are marked.}
         \label{FigExample}
   \end{figure}

For line intensity measurements, we simply integrated the line signal within a given spectral range and subtracted the background measured in the region around the given line. This approach is fast and straightforward for well-separated lines, as is the case for H$_\alpha$ and Si~II~-~2. Since both lines are faint in meteor spectra (see Figure \ref{FigExample}), saturation is not a problem for these measurements. We also estimated line intensity uncertainty based on measuring the standard deviation of the signal in a region near the line of interest, avoiding other emission features. When integrating line intensities over their spectral profiles, we propagated this background uncertainty across the integration bandwidth. For a line integrated over $n$ pixels, the statistical uncertainty scales as $\sigma_{\text{line}} = \sigma_{\text{bg}} \times \sqrt{n}$.

\section{Results}\label{results}
Many of the $1100$ spectra in our database showed just one or a few spectral lines. Only $31$ meteors showed resolved Si~II~-~2 doublet radiation at $634.71$~nm and $637.14$~nm, and $22$ of these spectra contained an H$_\alpha$ line at $656.28$~nm. All spectra that showed the H$_\alpha$ line also showed the Si~II~-~2 line. We can conclude that in digital photographic spectra of centimetre-sized meteoroids, the detection of neutral hydrogen and ionised silicon is relatively rare. This can be caused by several factors related to the conditions required for these emissions to be observable:

1. Both lines originate from the high-temperature component, which requires sufficient kinetic energy to be generated in the shock wave. In our studied sample, the slowest meteor with a detectable Si~II~-~2 line had a velocity of $28.48$~km/s, consistent with the previously observed minimal velocities needed for high-temperature component formation. Only about $42\%$of meteors in our spectral database—from more than 500 with computed velocities—have velocities exceeding this threshold.

2. The Si~II~-~2 and H$_\alpha$ emissions are intrinsically faint compared to dominant lines such as Na, Mg, Fe, and Ca~II (see Figure~\ref{FigExample}).

3. SDAFOs are optimised for spectra of bright fireballs. The faintest fireball with a visible H$_\alpha$ line had the magnitude of -10.

These observational constraints mean that the limited number of meteors in our sample may not necessarily reflect the actual abundances of hydrogen or silicon in asteroidal or cometary material, but rather the detectability of their spectral signatures under specific entry conditions.

\subsection{Relationship between H$_\alpha$ and Si~II~-~2 line intensities}

\begin{figure}
   \centering
   \includegraphics[width=\hsize]{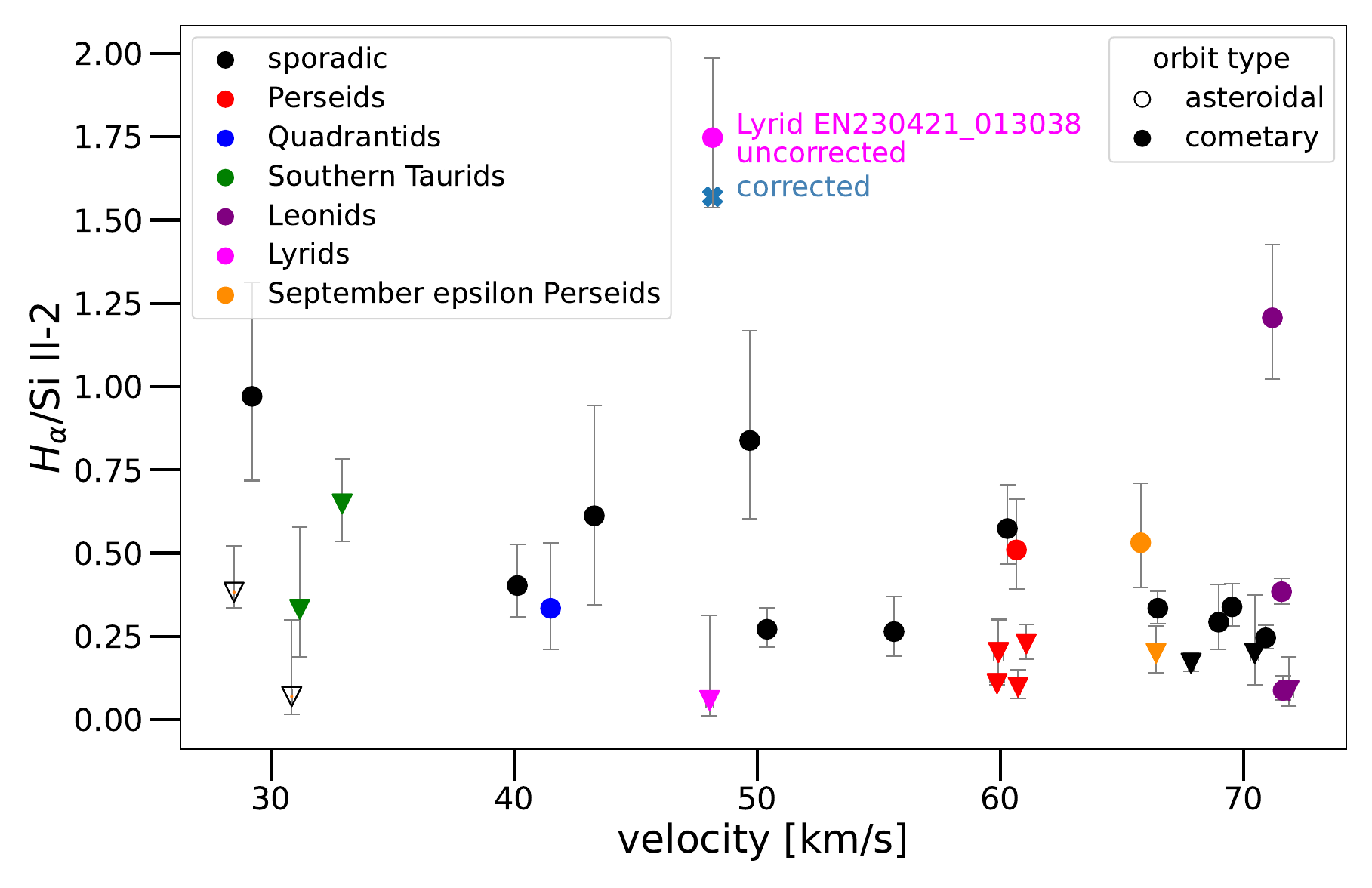}
      \caption{Ratio of the integrated intensities in the H$_\alpha$ region at $656$~nm to the silicon doublet Si~II~-~2 at $634$--$637$~nm as a function of meteor velocity. The different showers are marked in different colours. The circles mark the cases with clearly detectable hydrogen and silicon lines. The triangles mark the upper limits of the ratio when the hydrogen line was either weak, at the noise level, or invisible and only the level of noise was measured. The line ratio corrected for afterglow contamination in one Lyrid meteor is shown as a blue cross (see Sect. \ref{afterglow})}.
         \label{FigHSiVel}
   \end{figure}

When comparing integrated intensities of spectral lines, the ratio of the H$_\alpha$ line to the silicon Si~II~-~2 doublet shows no clear correlation with meteor velocity (Figure \ref{FigHSiVel}). This shows no significant dependence of the high-temperature component line intensity ratios on the meteoroid velocity, allowing us to study elemental abundances without velocity-related corrections. The average H$_\alpha$/Si~II~-~2 value is $0.41$ with a median of $0.33$. The average H$_\alpha$/Si~II~-~2 value for cases with clearly observable H$_\alpha$ is $0.55$ with a median of $0.39$. Several outliers exhibited significantly higher H$_\alpha$/Si~II~-~2 values, indicating potentially elevated hydrogen abundance. When excluding outliers with ratios above $0.75$, the average ratio decreases to $0.29$ with a median of $0.27$.

\subsubsection{Relationship with magnitude and photometric mass} 

\begin{figure}[htb]
\centering
   \begin{subfigure}{0.5\textwidth}
    \includegraphics[width=\hsize]{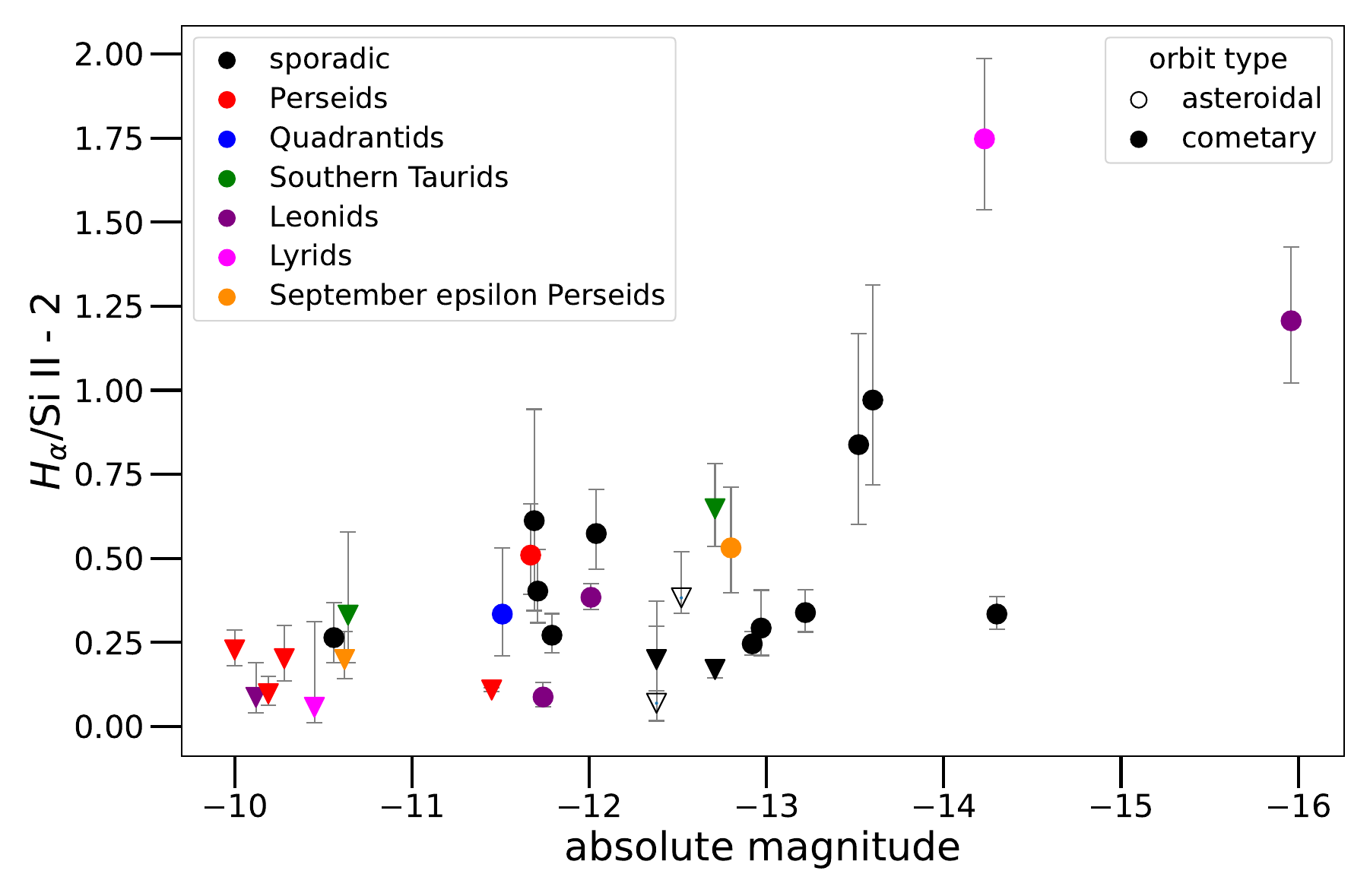} 
    \caption{}
    \label{FigPHMass_a}
  \end{subfigure}
   \begin{subfigure}{0.5\textwidth}
    \includegraphics[width=\hsize]{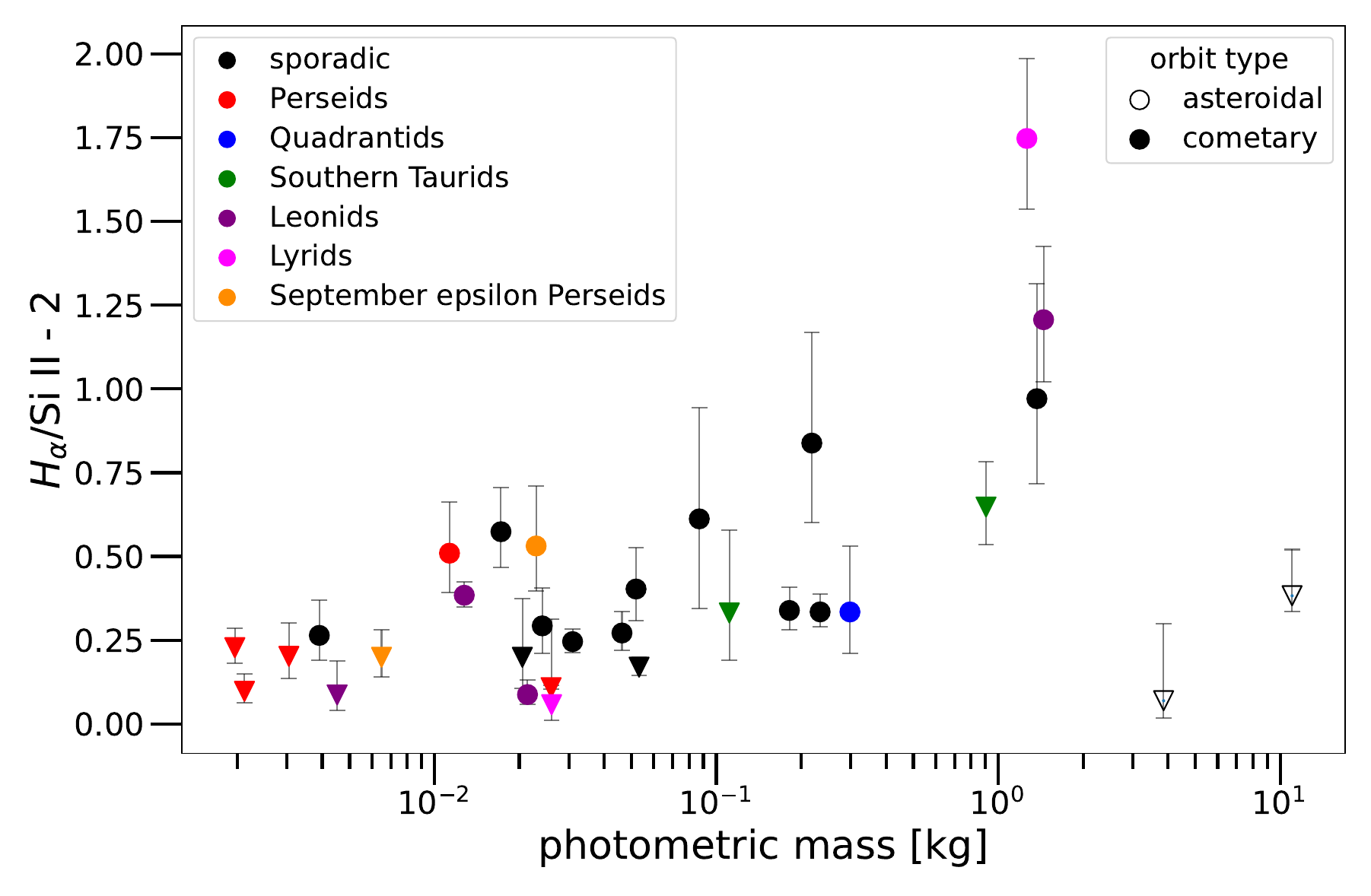} 
    \caption{}
    \label{FigPHMass_b}
  \end{subfigure}
  \caption{Ratio of the integrated intensities in the H$_\alpha$ region at $656$~nm to the silicon doublet Si~II~-~2 at $634$--$637$~nm. \textit{(a)}  As a function of absolute magnitude of the meteor. \textit{(b)} As a function of photometric mass of the meteoroid.}
     \label{FigPHMass}
\end{figure}

Figure \ref{FigPHMass} illustrates the relationship between the H$_\alpha$/Si~II~-~2 value and both absolute magnitude and photometric mass. For cometary meteoroids, the ratio remains relatively low (generally below $0.75$) up to a magnitude of approximately $-13.5$ or a photometric mass of $\sim100$~g. Beyond this threshold, the ratio increases substantially with increasing brightness and mass, with the trend becoming more pronounced for the largest meteoroids.

This observation is particularly significant as it suggests a potential mass dependence of volatile preservation in cometary material. The photometric mass calculation depends on the assumed luminous efficiency, which varies with velocity, composition, and atmospheric conditions. While the exact values of photometric mass carry significant uncertainty, the relative mass ordering is more reliable, supporting the observed trend.

Notably, the two asteroidal meteoroids (with Tisserand parameters $T_J > 3$ or aphelion $Q < 4.5$ AU) in our sample had high photometric masses but exhibited a low H$_\alpha$/Si~II~-~2 values, clearly diverging from the trend observed in cometary meteoroids. This suggests a fundamental compositional difference between asteroidal and cometary materials, with asteroidal meteoroids containing lower relative hydrogen abundances even at larger sizes.

\begin{figure}
  \centering
  \includegraphics[width=\hsize]{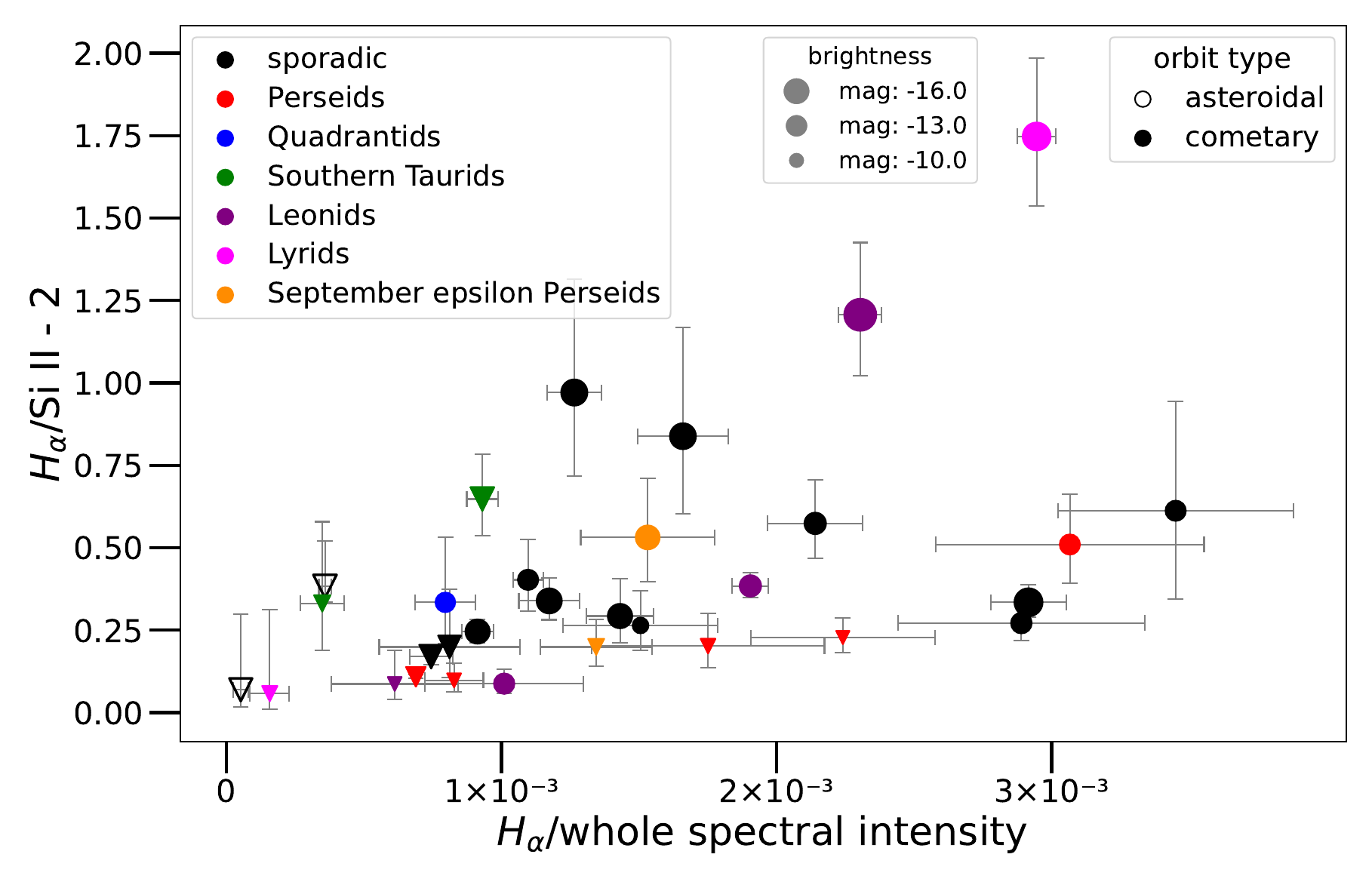}
  \caption{Ratio of the integrated intensities at H$_\alpha$ to the silicon doublet Si~II~-~2 as a function of the ratio of the intensity at H$_\alpha$ to the intensity of the whole observed spectrum. The size of each point corresponds to the absolute magnitude of the meteor.}
  \label{FigHSiWhole}
\end{figure}

Figure \ref{FigHSiWhole} presents the ratio of the H$_\alpha$ to the Si~II~-~2 intensities versus the ratio of the H$_\alpha$ to the total spectrum intensity. This provides insight into how hydrogen and silicon emissions relate to the overall spectral energy distribution. The whole spectral intensity was integrated across the $380$--$800$~nm range, with uncertainties derived from the standard deviation in line-free regions ($660$--$720$~nm). H$_\alpha$ emission constitutes a minor component of the total spectral energy, typically less than $0.3\%$ of the integrated intensity. For most meteors, H$_\alpha$ and Si~II~-~2 intensities increase mutually relative to the total spectrum, resulting in relatively consistent H$_\alpha$/Si~II~-~2 values across a range of brightnesses. However, several meteors deviate from this pattern, namely one Lyrid, one bright Leonid, and two sporadic meteors. Figure \ref{FigHSiWhole} also reveals a trend related to meteor magnitude, indicated by point sizes in the plot. Brighter meteors (larger symbols) tended to exhibit fainter relative Si~II~-~2 intensity and higher relative H$_\alpha$ intensity. This finding is consistent with the relationship between the H$_\alpha$/Si~II~-~2 value and photometric mass illustrated in Figure \ref{FigPHMass}, reinforcing the apparent correlation between meteoroid size and hydrogen abundance.

\subsubsection{H$_\alpha$ and Si II - 2 in shower meteors}   

Our sample includes both sporadic meteors and members of established meteor showers: five Perseids, four Leonids, two Southern Taurids, two Lyrids, two September $\epsilon$-Perseids, and a single representative from the Quadrantid shower. This diversity allows us to examine potential compositional differences between meteoroid streams.

Hydrogen is frequently detected in the Perseids and Leonids (e.g. \cite{Matlovic2022}).  \cite{Matlovic2022}  reported one case each in the Geminids and Orionids, but no bright Geminid or Orionid spectrum is contained in our sample. Both studies, however, detected hydrogen in Taurids—one Northern in \cite{Matlovic2022} and two in Southern (this work).

Given that Leonid masses varied from $4.5 \times 10^{-3}$ kg to $1.5$ kg, the meteors exhibited the most pronounced variability in their H/Si values. Their H$_\alpha$ emission ranges from undetectable to relatively strong, H$_\alpha$ intensity relative to the whole spectrum varied by a factor of $4$. 

Perseid meteors displayed a more apparently consistent H$_\alpha$/Si~II~-~2 value, with four samples showing low H$_\alpha$ emission and only one meteor having H$_\alpha$ clearly detectable above the noise. This can again be clearly explained by the relatively low masses of all Perseids. The relative intensity of H$_\alpha$ to the whole spectrum varied up to a factor of $4$, similar to the Leonids.

Two Lyrid meteors in our sample displayed two different H$_\alpha$/Si~II~-~2 values. While the very bright Lyrid showed the highest relative hydrogen intensity, the second, fainter Lyrid showed a typical H$_\alpha$/Si~II~-~2 value.

The two Southern Taurid meteors, associated with comet 2P/Encke, showed relatively low intensities of both H$_\alpha$ and Si~II~-~2 lines, partially attributable to their lower entry velocities (approximately $30$~km/s). This velocity is near the threshold for high-temperature component formation, limiting the excitation of these emission lines.

The Quadrantid and September $\epsilon$-Perseid meteors in our sample exhibited typical H$_\alpha$/Si~II~-~2 consistent with the main population trend. Even though both showers have very different velocities (see Figure~\ref{FigHSiVel}).

Figure \ref{FigPHMass} shows that most of the differences between individual shower members can be explained by the magnitude or mass dependence of the H$_\alpha$/Si~II~-~2 value with no suggestion of any kind of deviation for any of the studied showers.

\subsubsection{Orbital characteristics}

Most meteoroids in our sample have cometary origins, consistent with the requirements for high-temperature component formation. The two Southern Taurids are classified as ecliptic according to \cite{borovicka2005}, with comet 2P/Encke as their parent body. The remaining meteoroids have orbits typical of Halley-type comets.

We classified orbits as asteroidal if they had a Tisserand parameter $T_J > 3$ or aphelion $Q < 4.5$~AU. Only two meteoroids outside of Taurids (classified as ecliptic) in our sample met these criteria (see Figure \ref{FigHSiVel}). Both showed weak or barely detectable hydrogen emissions. Moreover, their spectra in the H-Si region contained multiple low-temperature component lines, contrasting with cometary meteor spectra where these lines were absent or sparse (see Figure \ref{FigAST}).

The predominance of cometary orbits in our sample is expected for two reasons. First, asteroidal meteoroids typically contain lower abundances of hydrogen-bearing materials, including water, compared to cometary bodies. Second, asteroidal meteors generally have entry velocities below $30$~km/s, insufficient to generate the high-temperature component necessary for H$_\alpha$ and Si~II~-~2 emissions.

Four meteoroid orbits in our sample had perihelia below $0.4$~AU; two of them are Southern Taurids. All exhibiting weak H$_\alpha$ emissions. This observation confirms the possible depletion of hydrogen in meteoroids with low perihelia reported by \cite{Matlovic2022}, likely resulting from increased solar heating and consequent volatile loss.

\subsection{Analysis of other high-temperature component lines}

In addition to H$_\alpha$ and Si~II~-~2, we measured lines of ionised magnesium (Mg~II--4 at $448.12$~nm and Mg~II--8 at $789.64$~nm) and the oxygen triplet at $777$~nm. The magnesium lines present specific challenges: Mg~II--4 overlaps with other meteor lines, while Mg~II--8 coincides with second-order spectrum features (see Figure \ref{FigExample}). Thus, measured values must be taken very carefully, with the high uncertainty of the measurements in mind. The oxygen triplet primarily originates from the atmosphere \citep{Vojacek2022}.

Figure \ref{fig:MgOWholeEnergy} shows the ratios of these additional high-temperature lines to hydrogen, presented in the same format as the Si~II~-~2 analysis in Figure \ref{FigHSiWhole}. As expected, most high-temperature lines showed similar behavior relative to hydrogen, with most spectra exhibiting consistent ratios except for several outliers showing elevated H$_\alpha$ emission.

Interestingly, the meteors that exhibited unusually high H$_\alpha$/Mg~II values were not always the same as those showing high H$_\alpha$/Si~II~-~2 values, suggesting element-specific variations rather than a uniform enrichment or depletion of hydrogen. This complexity highlights the compositional heterogeneity among meteoroids.

The oxygen line at $777$~nm is well-separated and particularly bright in fast meteors. Only two meteors—one Lyrid and one sporadic meteor—showed exceptionally high H$_\alpha$/O~I values. Both also exhibited high H$_\alpha$/Si~II~-~2 values. This suggests a high abundance of hydrogen for these meteoroids. Other meteors with elevated H$_\alpha$/Si~II~-~2 values showed normal H$_\alpha$/O~I values, suggesting they may have either anomalously low silicon or high oxygen abundances rather than universally elevated hydrogen.

\subsection{Relative atomic abundances}

\subsubsection{Neutral hydrogen and ionised silicon} 
To compute relative abundances of hydrogen and silicon atoms, we need to assume optically thin plasma and local thermodynamic equilibrium (LTE). If the radiation were optically thick, the studied lines would approach the intensity of H and K Ca~II lines (by far the brightest lines of the high-temperature component), which was not the case in any of the studied spectrum.
First, we estimated the relative ratio of radiating atoms of neutral hydrogen and ionised silicon using this equation, following the procedure in  \cite{Borovicka1993a}:

    \begin{equation}
   I = N \cdot \frac{e^2 h}{2 * m_e \varepsilon_0 U} \cdot \frac{g_1 f}{\lambda^3} \cdot e^{-E_2 / k_B T}.
\label{atomNum}
   \end{equation}

Here
\begin{itemize}
  \item \( N \) is the total number of atoms of the element in question;
  \item the constants \( h \), \( k_B \), \(\varepsilon_0 \), and  \( e \) are respectively Planck's constant,  Boltzmann’s constant, vacuum permittivity, and elementary charge;
  \item \( m_e \) is the electron mass;
  \item \( U \) is the partition function;
  \item \( g_1 \) is the statistical weight of the lower energy level;
  \item \( f \) is the oscillator strength of the transition, which quantifies the probability of the radiative transition;
  \item \( \lambda \) is the wavelength of the transition;
  \item \( E_2 \) is the excitation energy of the upper energy level;
  \item \( T \) is the absolute temperature of the gas.
\end{itemize}

The resulting line intensity units are $[I] = \frac {W}{ ster}$. This equation represents the radiation from a single atom multiplied by the total number of atoms of the given element in the plasma. We note that this form of the equation uses the SI unit system. In the CGS system, the term  $\frac{e^2 h}{2*  m \varepsilon_0 c}$ becomes  $\frac{2 \pi h e^2}{m_e}$ due to the different physical interpretation of the elementary charge: statcoulomb in CGS versus coulomb in SI. The only unknown quantity is the temperature $T$. Since we did not observe enough well-separated lines of one element in the high-temperature component, we were not able to estimate the temperature of the plasma. We adopted the result from previous work on high-dispersion spectra of the Leonids and Perseids in \cite{Borovicka2004} $T = 10500$~K. The result of the computed atom number ratios of neutral hydrogen H I to ionised silicon Si II for each meteor can be seen in Table \ref{tab:data} and also in Figure \ref{FigHI_SiII_atomNumberRatio}.

To test the  temperature sensitivity, we also estimated the ratio of neutral H atoms to ionised Si atoms at $10000$~K and $11000$~K. This range is based on the results for the high-temperature component in the work of \cite{Borovicka2004}. In Figure \ref{FigHI_SiII_atomNumberRatio}, the lower H\,I / Si\,II atomic ratio with higher temperature can be seen. This is primarily driven by the higher excitation energy of hydrogen compared to silicon ($12.09$~eV for H~I; $10.07$~eV for Si~II) in the Boltzmann distribution. Higher excitation energy means that, at a given temperature, fewer atoms are in the corresponding excited state. However, as the temperature increases, the population of high-energy levels grows more rapidly than that of low-energy levels, due to the exponential dependence in the Boltzmann distribution. Therefore, the intensity of high-excitation lines, for example  H$\alpha$, increases faster with temperature than low-excitation lines such as Si~II. Consequently, if we assume a higher temperature, fewer hydrogen atoms are needed to explain the observed intensity of the H$\alpha$ line. Another temperature dependent variable in   Equation \ref{atomNum} is the partition function. Although the partition function $U(T)$ depends on the temperature, for H\,I it is essentially flat, while Si\,II increases only slightly in the given temperature range.  Thus, the Boltzmann factor plays a dominant role in the temperature dependence of the number of atoms.  
However, we can see in the Figure \ref{FigHI_SiII_atomNumberRatio} that this systematic variation across the 1000K temperature range remains well within uncertainty of the H~I/Si~II measurement.

\subsubsection{Relative hydrogen and silicon atomic abundance} 

When comparing relative abundances of elements, a correction for ionisation must be performed since, for given plasma conditions, some atoms can be in neutral form and some can be in ionised form.
Using Saha equation, we can compute ionisation correction:
\begin{equation}
\frac{N_2}{N_1} = \frac{2 U_2}{U_1} \left( \frac{2 \pi m_e k_B T}{h^2} \right)^{3/2} \cdot \frac{1}{n_e} \cdot  e^{-E_i / k_B T}
\label{Saha}
\end{equation}

\begin{figure}[htb]
\centering
   \begin{subfigure}{0.5\textwidth}
    \includegraphics[width=\hsize]{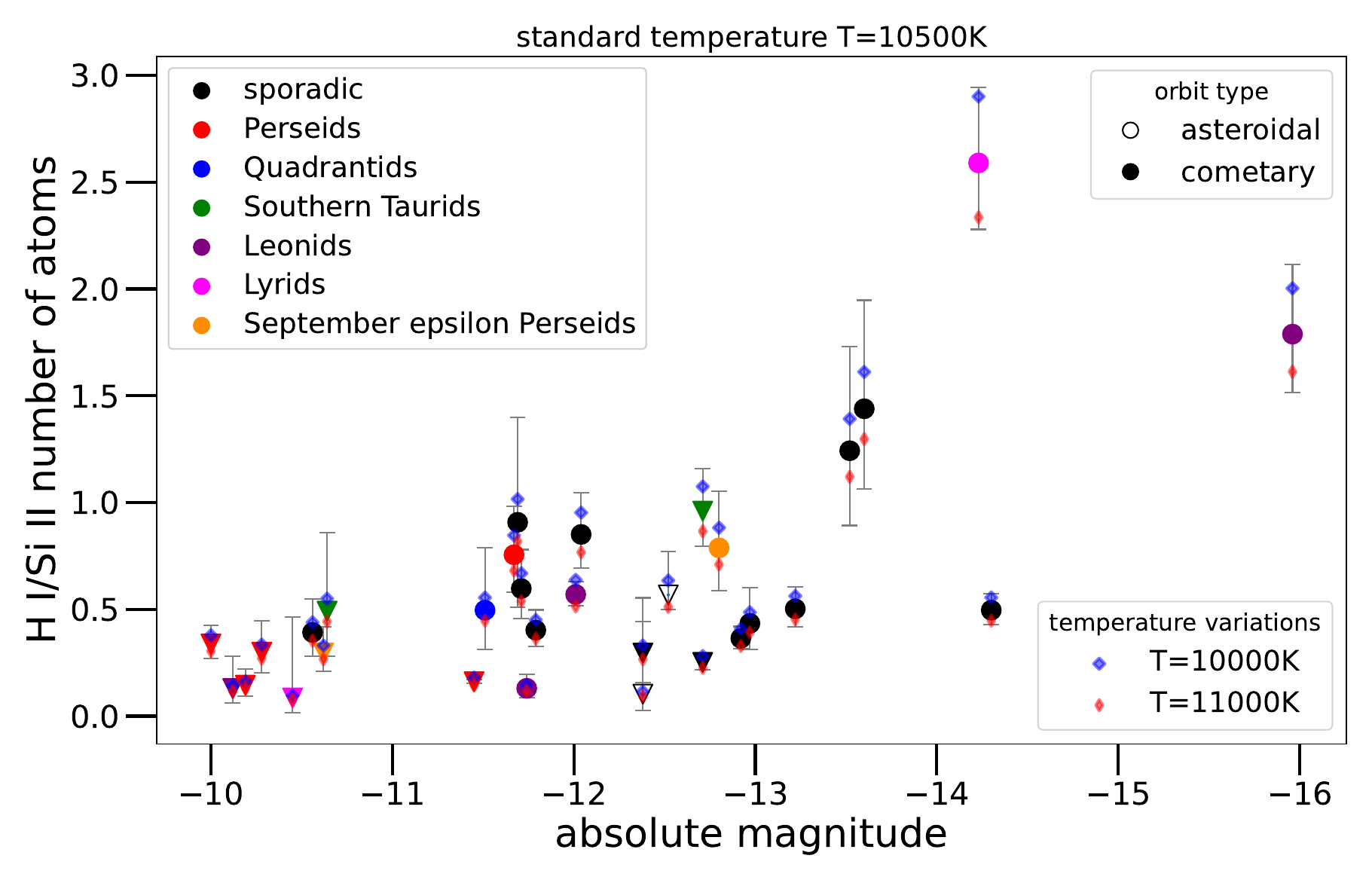} 
    \caption{}
    \label{FigHI_SiII_atomNumberRatio}
  \end{subfigure}
   \begin{subfigure}{0.5\textwidth}
    \includegraphics[width=\hsize]{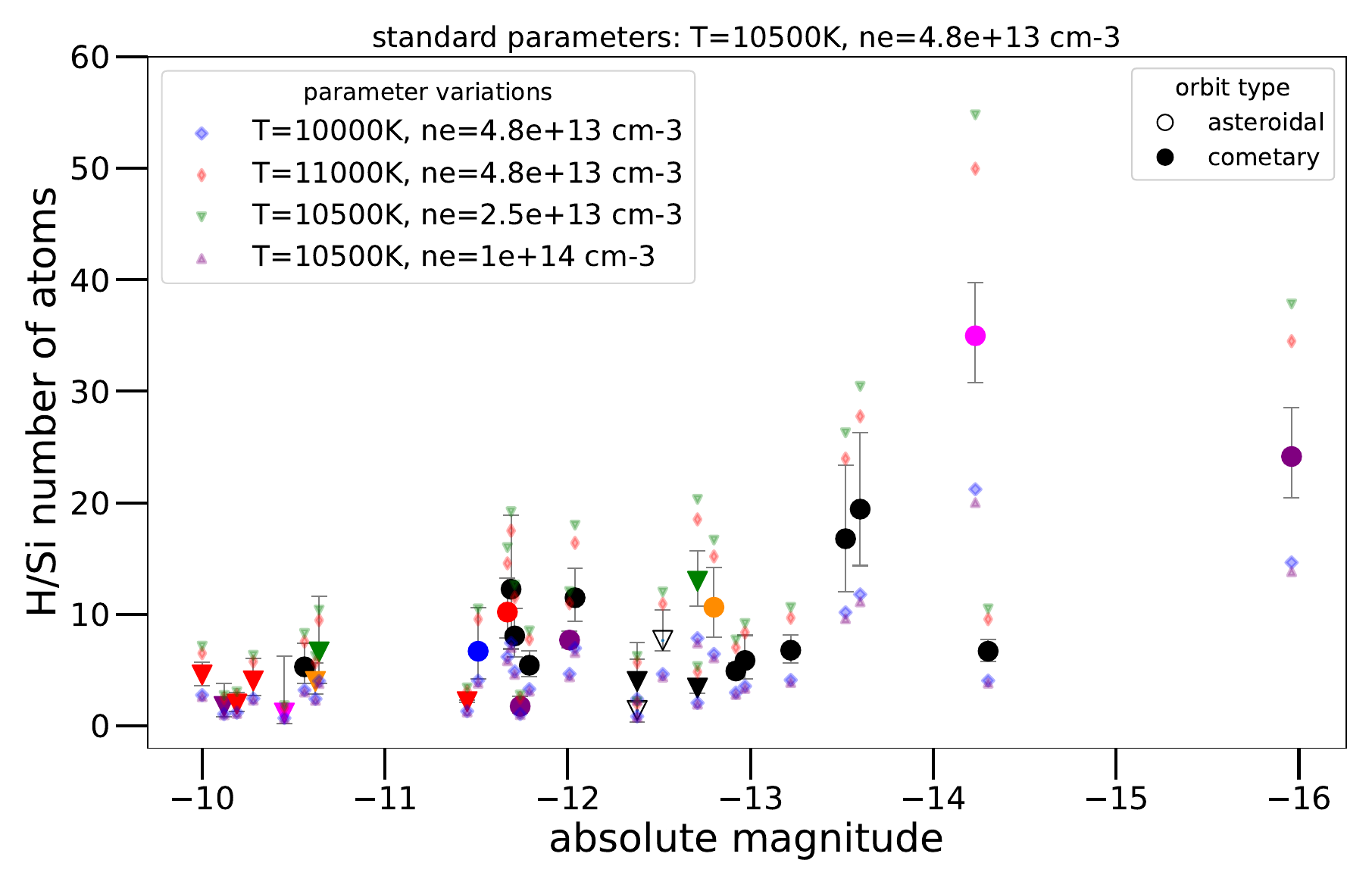}
    \caption{}
    \label{FigH_Si_atomNumberRatio}
  \end{subfigure}
  \caption{Ratios of the number of atoms. \textit{(a)} For neutral hydrogen and ionised silicon for standard temperature; two temperature variations are also shown. \textit{(b)} For hydrogen and silicon (after ionisation correction) for standard temperature and electron density. The ratios for the temperature and electron density variations in plasma are also shown (see left inset). }
     \label{atomNumberRatio}
\end{figure}

\begin{figure}[htb]
\centering
    \includegraphics[width=\hsize]{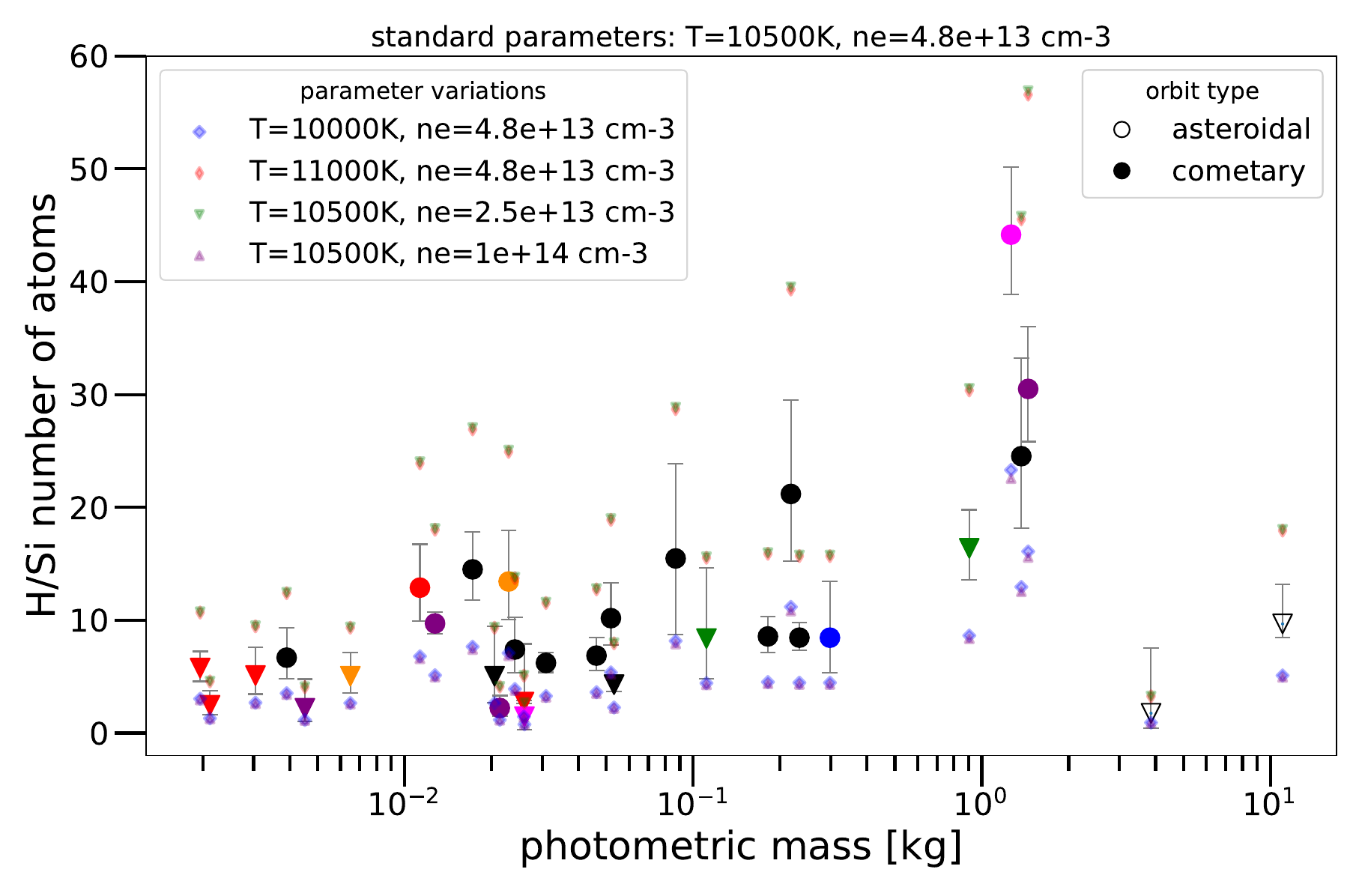}
  \caption{Same as Figure \ref{FigH_Si_atomNumberRatio}, but for meteoroid photometric mass. } 
     \label{atomNumberRatioPhMass}
\end{figure}

Where \(N_2\) is number of atoms in upper state and  \(N_1\) is number of atoms in lower state. \(U_1\) and \(U_2\) are partition functions for given atomic state, and \(n_e\) is electron density of the plasma. We assumed the same temperature as in Equation \ref{atomNum}. Since we studied low-resolution spectra of faint lines and we did not observe well separated lines of one element from both low-temperature and high-temperature components, we were not able to estimate temperature and electron density from our observations and we the assumed same electron density for all meteors. This electron density was also taken from \cite{Borovicka2004} $n_e = 4.8 \times 10^{13} \, \text{cm}^{-3}$. Results can be seen in Table~\ref{tab:data} , in Figure~\ref{FigH_Si_atomNumberRatio} for meteor magnitude and in Figure~\ref{atomNumberRatioPhMass} for meteoroid photometric mass.

Additionally, we performed a sensitivity analysis using the Saha equation to assess how ionisation corrections affect the atomic H~I/Si~II value under varying plasma conditions. The analysis considered both temperature variations ($10000-11000$~K) and electron density variations ($2.5 \times 10^{13}$ to $1.0 \times 10^{14}$ cm$^{-3}$), which bracket typical meteor plasma conditions of high-temperature component, based on  \cite{Borovicka2004}. Our results revealed that both hydrogen and silicon exhibited high ionisation fractions under these conditions, with silicon being nearly completely ionised (>99.9$\%$), while hydrogen ionisation varies from $\approx77\%$ to $\approx98.5\%$.  Moreover, silicon under the assumed plasma conditions consists of approximately 80\% singly ionised atoms Si~II and about 20\% doubly ionised silicon Si~III. Both of these ionisation states were considered when computing the total number of Si atoms. For details see Tables \ref{tab:hydrogen_compact} and \ref{tab:silicon_compact}.

In the Figure \ref{FigH_Si_atomNumberRatio} there are plotted four other variations of H/Si atom number ratio for combinations of temperatures 10000~K and 11000~K and electron densities $2.5 \times 10^{13}$ and $1.0 \times 10^{14}$ cm$^{-3}$. We can see that there can be difference up to factor of $2.7$, usually higher than uncertainty of the line intensity measurements. This high variability is given by high sensitivity of hydrogen ionisation in this ranges of plasma parameters. The lower the temperature is and higher the electron density is the smaller is the hydrogen ionisation and we receive lower number of all hydrogen atoms. The overall ratio of all hydrogen atoms to silicon atoms is then lower. This is to some extent compensated by the fact that we observe only singly ionised silicon, but in the plasma there is significant amount of doubly ionised silicon Si~III. The lower the temperature is and higher the electron density is the higher is the percentage of Si~II in all silicon and thus smaller number of atoms of all silicon.

Assuming the same plasma parameters for all meteors in the sample means simply different scaling of measured line ratios with no effect on dependence on meteor parameters such as magnitude photometric mass. Theoretically, brighter meteors are more likely to penetrate deeper into the atmosphere, where the electron density should be higher. We investigated if the magnitude of meteors significantly depended on the altitude at which given spectrum was scanned. Figure \ref{Fig:MagScan} shows that there is no specific trend. Thus, we decided to assume the same electron density for all spectra.

We divided our sample into four mass bins to demonstrate the mass dependence of H/Si more clearly. Table~\ref{tab:hsi_comparison} shows the average H/Si value for each mass bin with its associated uncertainties. These uncertainties combine measurement errors and statistical uncertainties. We also present a nominal combined value, for which we adopted the H/Si value derived from assumed plasma parameters. The uncertainty for this combined value incorporates both the uncertainties presented for this ratio and the uncertainty arising from plasma parameter variations. The variation uncertainty was calculated as half the difference between the maximum and minimum H/Si values obtained across all parameter variations. The combined values with their uncertainties are also presented in Figure~\ref{FigH_Si_Comparison}. For smallest mass bin ($<$~0.01~kg) the H/Si value is $3.6 \pm 1.5$, but for largest meteoroids ($>$~0.5~kg) the ratio is $23 \pm 10$.

\section{Discussion}\label{discussion}

\subsection{Origin of higher hydrogen abundance}

As shown in Figure \ref{FigHSiVel}, the relative radiation of the H$_\alpha$ line to the ionised silicon Si~II~-~2 doublet varies significantly among meteors. While we found no strong correlation between H/Si values and orbital parameters, except for the absence of these lines in low-perihelion meteoroids, we observed a clear trend of increasing H~I--I/Si~II--2 value (and consequently increasing H/Si value) with increasing absolute magnitude and photometric mass for cometary meteoroids (see Figure \ref{FigPHMass}). This suggests that larger cometary meteoroids may better preserve their primordial volatile content, potentially due to reduced surface-to-volume ratios limiting volatile loss through space weathering processes.

A plausible explanation for this mass-dependent pattern lies in the heating history of meteoroids during their orbital evolution. Smaller meteoroids have higher surface-to-volume ratios, exposing proportionally more material to solar heating, cosmic radiation, and space weathering. These processes preferentially deplete hydrogen from smaller bodies, while larger meteoroids retain more of their original volatile material due to their lower surface-to-volume ratios and consequently reduced effects of surface processes. Interestingly, recent laboratory work by \cite{Libourel2021} demonstrated that thermal cycling induces networks of cracks in meteorite samples, with enhanced crack formation in hydrated specimens. Sheet silicate interlayer water is progressively lost, giving rise to tensile stress by which hydrogen in water can be lost while silicon in silicates remains, potentially affecting H/Si values. It is important to note that they studied asteroidal meteorites containing terrestrial water, while in our study, most samples were of cometary origin.

While our data strongly suggest a correlation between photometric mass and H/Si value for cometary meteoroids, we must consider several alternative explanations before conclusively attributing this trend to preferential hydrogen preservation in larger bodies:

First, observational bias could potentially influence our results. Both Si~II~-~2 and H$_\alpha$ lines are faint in meteor spectra, and thus the mass dependence can reflect detection limitations. If Si~II~-~2 is brighter than H$_\alpha$ (as observed in most cases), and H$_\alpha$ is near the detection threshold, brighter meteors might naturally show higher H$_\alpha$/Si~II~-~2 values as hydrogen becomes more detectable. However, if this were merely an observational effect, we would expect a more linear growth in the ratio from the faintest to the brightest meteors, and the ratio would naturally not exceed a value of $1$. Yet we observe H$_\alpha$/Si~II~-~2 values larger than $1$ in the Lyrid and one Leonid, demonstrating that hydrogen emission can indeed be stronger than silicon emission. This indicates that the observed trend represents a genuine compositional effect rather than an observational artefact.

An alternative explanation for the observed mass-brightness relationship could involve differential optical thickness effects. If the plasma in brighter meteors were more optically thick for Si~II~-~2 compared to H$_\alpha$, we might observe a similar trend to that shown in Figure \ref{FigPHMass}. However, the high-temperature component of the plasma appears to be rather optically thin. This is evident from the high intensity of the H and K lines of Ca~II at approximately $390$~nm, while other high-temperature component lines, such as Si~II~-~2 and H$_\alpha$, remain relatively faint (see Figure \ref{FigExample}). If the medium were optically thick, the maxima of lines originating from the same component would tend to follow the Planck radiation profile, which is not observed in our spectra.

Another possibility is that our correlation reflects genuine compositional differences between larger and smaller cometary fragments. Heterogeneity within the cometary body, with more volatile-rich material preserved in larger pebbles, would produce similar observational results. This heterogeneity might originate from the comet formation process itself or from subsequent evolutionary processes affecting different-sized fragments differently. Whether the mass-dependent hydrogen-to-silicon ratio is a consequence of interplanetary environment influence on meteoroids, early Solar System evolution, or a combination of both, remains a question for further research.

We do not observe a clear dependence of the H/Si value on perihelion distance, though most meteoroids in our sample had perihelion distances larger than $0.6$~AU, where the effect of solar radiation on volatile content would be more limited. The few meteoroids with lower perihelia in our sample tended to have low intensities of both hydrogen and silicon, consistent with findings by \cite{Matlovic2022} showing hydrogen depletion in meteoroids with smaller perihelion distances.

\cite{Borovicka2004} reported large scatter in Na and H values for the Leonids. The Perseids showed rather low H/Si values, with only one meteor having H$\alpha$ well above the noise level. \cite{BorovickaJenniskens2000} found the ratio of the number of atoms of hydrogen to another refractory element Fe—H/Fe—in photographic Leonids to be larger than that observed in the Perseids by \cite{BorovickaBetlem1997}. While our measurements showed comparable H/Si for most Leonids to Perseid meteors, only the very bright Leonid exceeded all other members of both Leonids and Perseids in our sample. Comparison of hydrogen to silicon ratios for Southern Taurid meteors, associated with comet 2P/Encke, showed low intensities of H$\alpha$ lines, consistent with their lower entry velocities (approximately 30 km/s). The relatively low velocity means less kinetic energy is available to excite the high-temperature component, making the detection of both elements more challenging.
The bright Lyrid meteor (parent body comet C/1861 G1 Thatcher) displayed the highest relative intensity of hydrogen compared to silicon in our entire sample. In contrast, the second Lyrid showed a low H/Si value while also having a much fainter magnitude. Despite these individual variations among shower members, the observed compositional differences can be largely explained by the systematic mass dependence of the H/Si value, indicating that meteoroid size represents the primary controlling factor in hydrogen-to-silicon abundance patterns across different meteor streams.

Based on the previous discussion, we can examine and compare meteors from different showers that have similar photometric masses (see Figure  \ref{FigPHMass} or Figure \ref{atomNumberRatio}). If we categorise meteors into groups such as up to 0.01 kg, 0.01–0.5 kg, and above 0.5 kg, we find that significant differences in the H/Si value appear only in the group above 0.5 kg. In this mass range, the Southern Taurid shows a much lower H/Si value, while the Leonid exhibits a higher value, and the Lyrid displays the highest ratio among them. However, these are only individual representatives of their respective showers, so no definitive conclusions can be drawn about systematic differences between meteor showers. Additionally, in this higher mass range, the dependence of the H/Si value on mass becomes more pronounced. For meteors in the lower mass bins, no clear differences in H/Si value between different showers can be identified.

\subsection{Afterglow contamination and its impact on H/Si measurements}\label{afterglow}

While Si~II~-~2 at $634.71$--$637.14$~nm and H$_\alpha$ at $656.28$~nm are well separated from other spectral lines from the meteor head or shock wave, there is a Ca~I line at $657.46$~nm and an Fe~I line at $635.9$~nm originating in the meteor afterglow. These intercombination lines were observed in video spectra of Leonids by \cite{Jenniskens2004Hydrogen}, where the Ca~I line significantly exceeded the H$_\alpha$ line. Video spectroscopy has the advantage of easily resolving meteor afterglow from the meteor head temporally and spatially. Photographic cameras with long exposures, however, cannot resolve objects flying behind each other. Moreover, the spectral resolution of our SBK| cameras is insufficient to resolve H$_\alpha$ from the intercombination line of Ca~I.

To estimate whether the observed H$_\alpha$/Si~II~-~2 could be influenced by afterglow contamination, we searched for other well-separated afterglow lines in our meteor spectra. If any afterglow line is detected in a spectrum, it suggests favourable conditions for other intercombination lines, including Ca~I at $657.46$~nm and an Fe~I line at $635.9$~nm that could contribute to the measured H$_\alpha$ and Si~II~-~2 radiation. The Mg~I line at $457.1$~nm serves as a convenient indicator for afterglow presence. Other intercombination lines to consider include Fe~I~-~2 at $442.7$~nm and  Fe~I~-~1 lines at $511.0$~nm and $516.6$~nm.

\begin{figure*}[htb]
\centering
\begin{minipage}{0.7\textwidth}
    \centering
    \includegraphics[width=\textwidth]{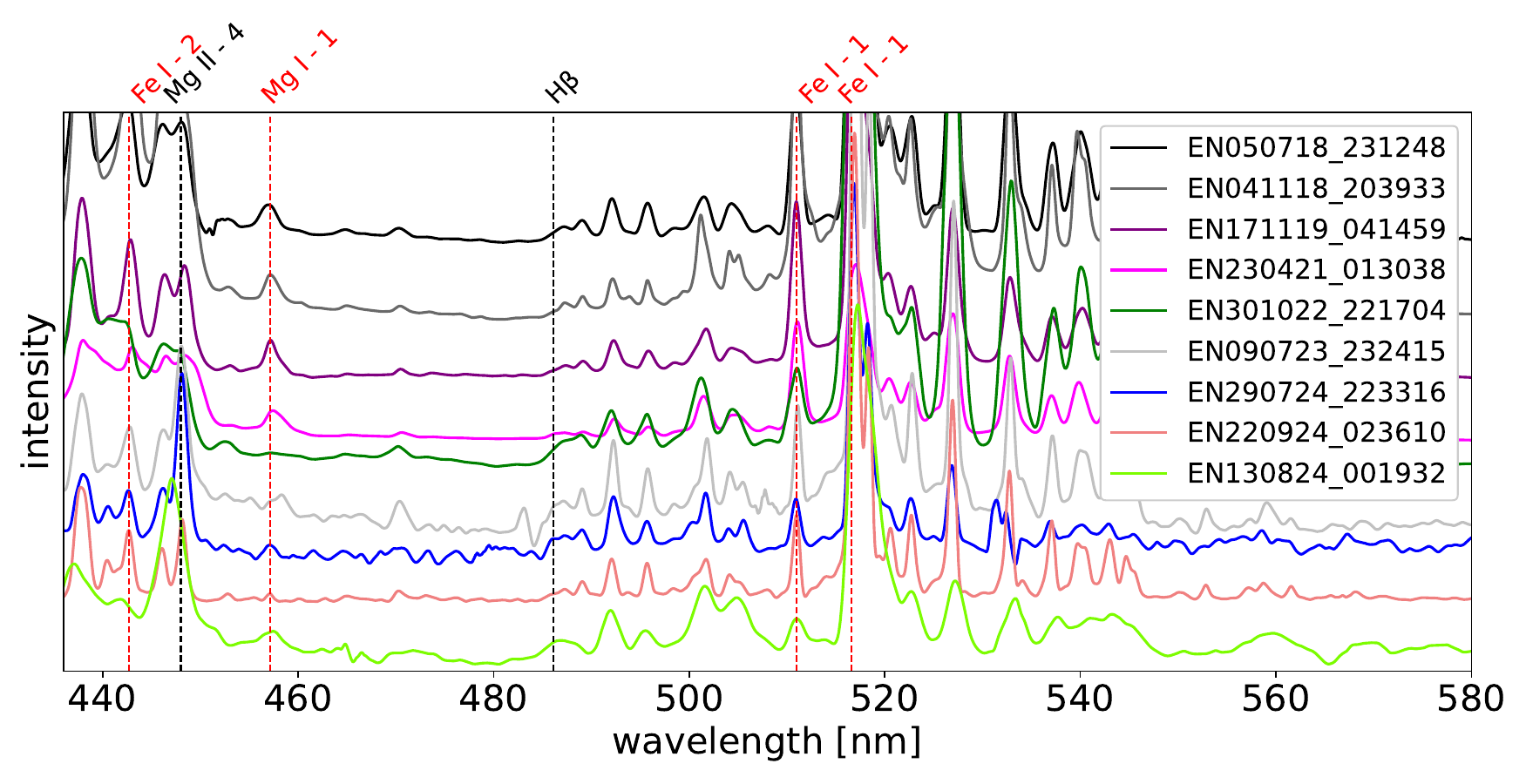}
\end{minipage}%
\begin{minipage}{0.25\textwidth}
    \raggedright
    \captionof{figure}{Spectral region between $440$~nm and $580$~nm of meteors with intercombination Mg~I line. The spectral intensities are manually scaled and offset. The theoretical positions of the high-temperature lines are marked with black vertical dashed lines and the positions of the intercombination lines are marked with red vertical dashed lines.}
    \label{FigMgRecombVyber}
\end{minipage}
\end{figure*}

Figure \ref{FigMgRecombVyber} shows the spectral region $440$~nm–$580$~nm for meteors whose H$_\alpha$/Si~II~-~2 exceeded a value of $0.5$ in Figure \ref{FigHSiVel}.  While the Fe~I--1 intercombination line at $516.6$~nm overlaps with a strong magnesium line and multiple Fe~I lines from the meteor head, the Mg~I--1 intercombination line at $457$~nm is well separated and visible in several cases, indicating afterglow presence in these spectra. This suggests that for the Lyrid and the Leonid with high H/Si values, the actual hydrogen overabundance might be lower than measured, as afterglow lines may contribute to the signal at the H$_\alpha$ wavelength.

\begin{figure}
\begin{subfigure}{.5\textwidth}
  \centering
  \includegraphics[width=\hsize]{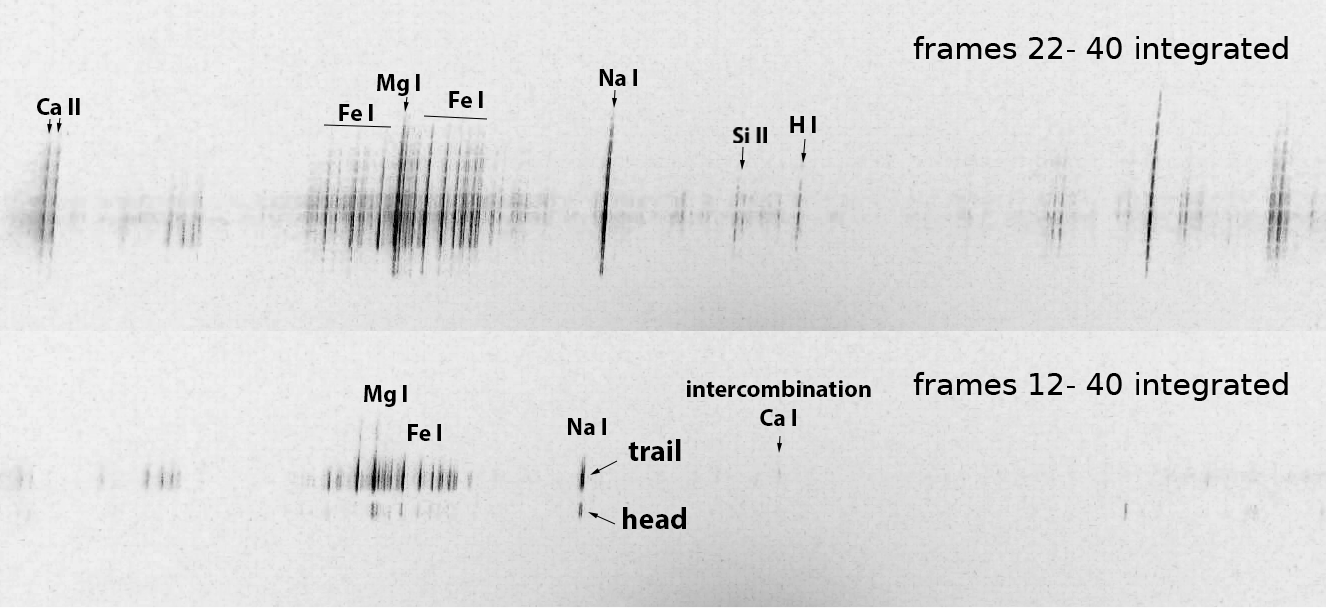}
 \caption{}
\label{LyrVideoFramesIntegrateda}
\end{subfigure}%
\\
\begin{subfigure}{.5\textwidth}
  \centering
  \includegraphics[width=\hsize]{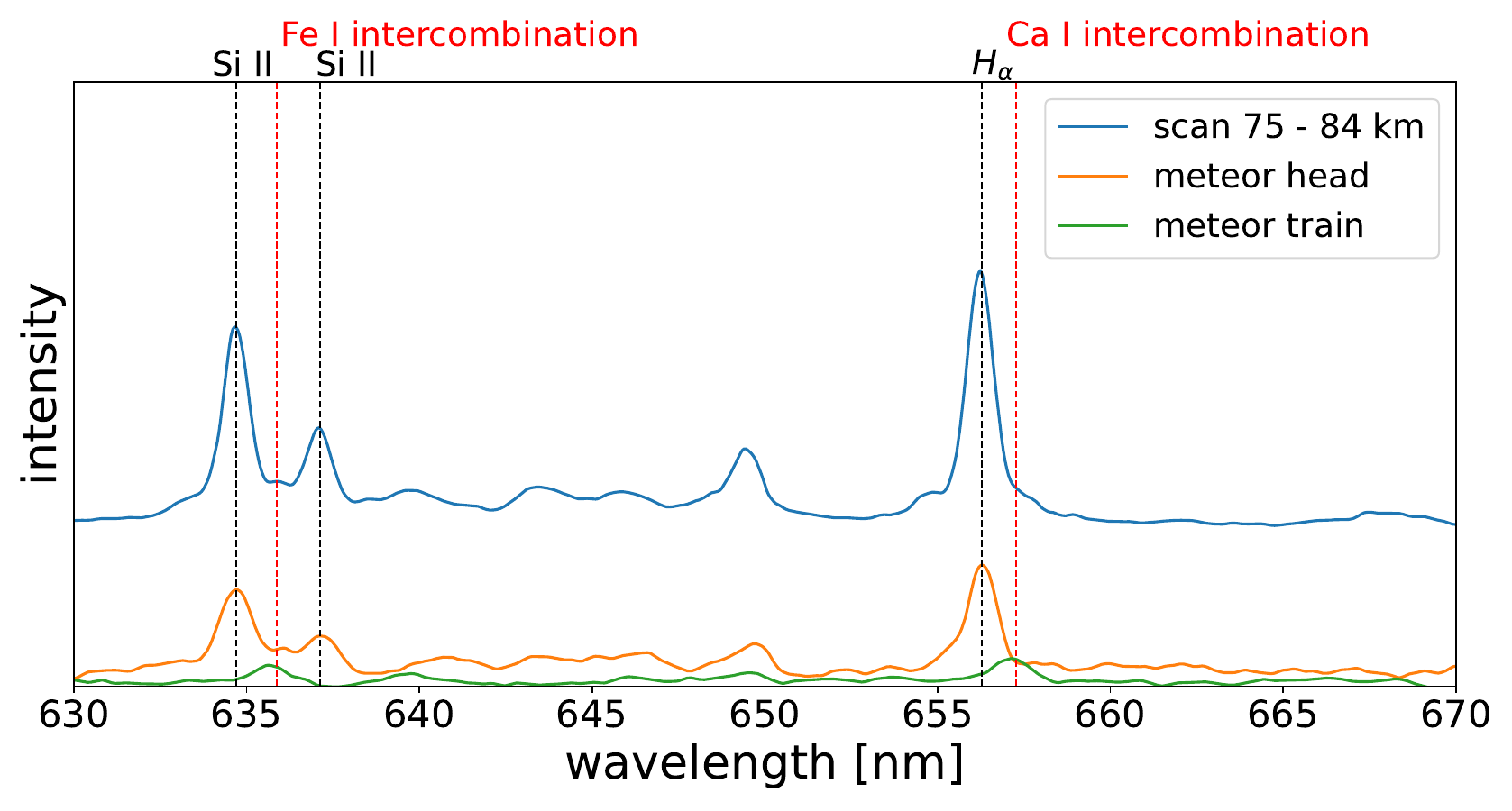}
\caption{  }
\label{LyrVideoFramesIntegratedb}

\end{subfigure}
  \caption{Integrated video frames of Lyrid meteor EN230421\_013038 with its spectrum captured by IP camera and the scanned part of the spectrum.  \textit{(a)} Integrated video frames. Upper panel: Whole meteor integrated. Bottom panel: Last images of the video with separated main meteor and meteor trail.    \textit{(b)} Scanned whole spectrum and spectra of meteor head and meteor train. }
  \label{LyrVideoFramesIntegrated}
\end{figure}

Fortunately, the brightest Lyrid fireball spectrum EN230421\_013038 was also captured by IP video cameras, which are part of the European Fireball Network. This camera is equipped with a spectral grating that allowed separation of meteor spectrum from the afterglow spectrum. The composite image can be seen in Figure \ref{LyrVideoFramesIntegrated}. In Figure \ref{LyrVideoFramesIntegrated}, the resolved meteor head and trail (afterglow) can be clearly distinguished. The video itself shows a moving meteor head followed by a non-moving spectrum of meteor afterglow after the meteor passed a given point. Scanned spectra reveal clear differences between meteor head and trail, with Si~II~-~2 doublet and strong H$_\alpha$ visible in the meteor head while intercombination lines of Ca~I and Fe~I appear in the meteor afterglow.

In contrast to the Leonid spectrum studied by \cite{Jenniskens2004Hydrogen}, this Lyrid spectrum shows stronger H$_\alpha$ compared to the intercombination Ca~I line. This demonstrates that even when the intercombination line Mg~I at $457$~nm is clearly visible (indicating relatively strong intercombination spectrum), the intercombination Ca~I line can represent just a fraction of the H$_\alpha$ intensity.

We can quantify the relative contributions of intercombination lines to the high-temperature lines of Si~II~-~2 and H$_\alpha$. The measured ratio of intercombination lines Ca~I/Fe~I was $1.44$. Since the IP cameras are not calibrated for absolute intensity, we computed relative intensities of H$_\alpha$ to Ca~I and Si~II~-~2 to Fe~I, which share very close wavelengths. The H$_\alpha$/Ca~I value was $3.6$, meaning the intercombination line present in the afterglow contributes about $22\%$ to the measured intensity at the H$_\alpha$ wavelength in unresolved spectra. The value of Si~II~-~2/Fe~I was $6.5$, indicating iron contributes about $13\%$ to the Si~II~-~2 intensity in unresolved observations.

We can use these ratios to estimate a correction to the H$_\alpha$/Si~II~-~2 value measured in our SBK observations. The corrected H$_\alpha$/Si~II~-~2 for the Lyrid meteor is shown in Figure \ref{FigHSiVel} and is marked as `Lyrid, corrected'. The correction lowered the H$_\alpha$/Si~II~-~2 value of the Lyrid spectrum from $1.75$ to $1.57$.

We must be careful not to apply this specific correction factor to all meteors in our sample. The strength of afterglow contamination likely varies with meteoroid mass, as Figure \ref{FigMgRecombVyber} suggests. Afterglow is present at the position of bright flares where a large amount of mass was ablated in a short time interval and the subsequent cooling of the plasma is slow \cite{BorovickaJenniskens2000}. For some meteors with high H/Si values we observed relatively strong Mg~I~-~1, Fe~I-~-2 and Fe~I~-~1 intercombination lines. This suggests that intercombination lines may also be contributing to the apparent hydrogen signal of other meteors. Without direct spectral separation similar we had for the Lyrid, we cannot precisely measure this effect.

Despite these uncertainties, several factors suggest that the observed correlation between H$_\alpha$/Si~II~-~2 value and photometric mass is real. Even after correction, the Lyrid meteor still shows a high H$_\alpha$/Si~II~-~2 compared to smaller meteors. Moreover, the trend appears across multiple meteor showers, which argues against afterglow contamination as the main cause, since afterglow characteristics differ between showers due to different entry velocities and compositions.

Despite afterglow correction, the Lyrid meteor maintains the highest H$_\alpha$/Si~II~-~2 value. The spatially separated Ca~I and H$_\alpha$ lines in IP camera observations ensures their isolation, with residual Ca~I contribution likely minimal. While the corrected H$_\alpha$/Si~II~-~2 value of 1.57 initially appears high compared to other meteoroids of similar mass, in general the H/Si value versus mass relationship shows considerable variation even at intermediate masses where we have the most data. This variation, caused by measurement uncertainties and real compositional differences, likely continues to higher masses. Considering this, the Lyrid's elevated ratio becomes less anomalous within the expected range of variation.

Future observations would benefit from spectroscopic systems that can better separate meteor head and afterglow emissions. Such measurements could provide shower-specific corrections that would improve the accuracy of the derived element abundances.

\subsection{Comparison of H/Si values across Solar System materials}

\begin{figure*}[htb]
\centering
\begin{minipage}{0.7\textwidth}
    \centering
    \includegraphics[width=\textwidth]{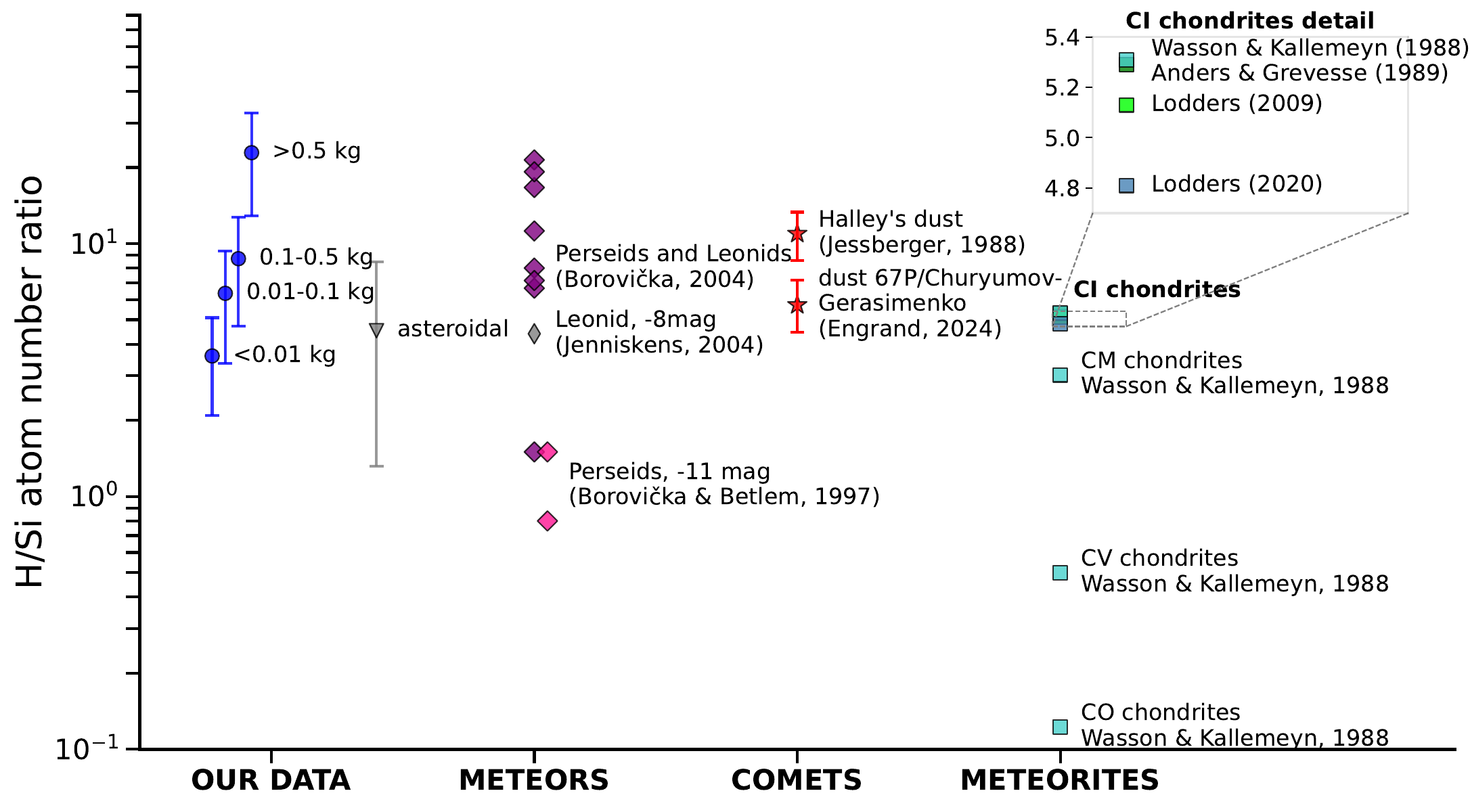}
\end{minipage}%
\begin{minipage}{0.25\textwidth}
    \raggedright
    \captionof{figure}{Comparison of measured H/Si atomic abundances with abundances from other meteor studies (\cite{BorovickaBetlem1997}, \cite{Borovicka2004}, \cite{Jenniskens2004Hydrogen}); with studies on cometary dust (\cite{Jessberger1988}, \cite{Engrand2024}); and with carbonaceous chondrites (\cite{Wasson1988}, \cite{AndersGrevesse1989}, \cite{Lodders2009}, and \cite{Lodders2020}.}
    \label{FigH_Si_Comparison}
\end{minipage}
\end{figure*}

\subsubsection{Previous meteor spectroscopic studies}

The comparison with previous meteor studies can be seen in Figure~\ref{FigH_Si_Comparison} and in Table~\ref{tab:hsi_comparison}. Only a limited number of meteor spectroscopic studies have reported H/Si values. \cite{BorovickaBetlem1997} studied two bright Perseid fireballs and found relatively low H/Si values comparable to some individual meteors in our sample including the Perseids, though their hydrogen measurements were limited to use the faint H$_\beta$ line due to wavelength sensitivity constraints. \cite{Borovicka2004} measured elemental abundances for five Perseid and four Leonid fireballs, reporting atomic H/Si values from $1.5$ to $21.4$ that span the range observed in our sample. \cite{Jenniskens2004Hydrogen} estimated a single H/Si value of $4.4$ for a video Leonid that falls within our observed range.

Leonid H/Si values exhibited the highest variability, ranging from $1.74$ to $24.2$, with this spread connected to the large photometric mass range of the Leonids observed. Perseids showed a narrower range of H/Si values. When comparing meteors of similar masses between showers, we find no significant systematic differences.

While \cite{BorovickaJenniskens2000} does not provide H/Si values, they reported H/Fe for Leonid meteors of approximately $10$-$20$ and compared this with Perseids in \cite{BorovickaBetlem1997} where the H/Fe value was approximately $2$-$6$. When comparing members of these shower with similar masses, we do not see significant differences of H/Si values (see Figure \ref{FigPHMass}).

\cite{Matlovic2022} reported generally higher H$_\alpha$/Na~I~-~1 values than we observed in our sample. While our maximum and median values were 0.17 and 0.06, respectively, about half of their spectra exceeded 0.2, with a maximum above 0.6.  According to personal communication with co-author Adriana Pisar\v c\' ikov\' a, the relatively strong hydrogen cases could be influenced by noisy, faint spectra with strong background, where manual fitting and imperfect subtraction could increase apparent signals and uncertainties.

\subsubsection{Cometary dust measurements}

Space mission data provide direct measurements of cometary dust composition for comparison. \cite{Jessberger1988} measured H/Si abundances in 1P/Halley dust as $10.95$ using the PUMA--1 mass spectrometer, while \cite{Engrand2024} derived H/Si values of $5.72$ for comet 67P/Churyumov--Gerasimenko from published H/C and C/Si measurements \citep{Bardyn2017, Isnard2019}. These space-based values exceed our computed H/Si for small meteoroids ($<0.01$ kg), with Halley's measurements also higher than our small-to-medium meteoroids ($<0.5$ kg, Figure~\ref{FigH_Si_Comparison} and Table~\ref{tab:hsi_comparison}). Our largest cometary meteoroids ($>0.5$ kg) showed H/Si = $22.88^{+5.05}_{-4.92}$ for default plasma parameters, twice Halley's dust ratio. However, H/Si values are highly sensitive to plasma conditions. Assuming 500K lower temperature or half the electron density yields H/Si $\approx 13-14$, consistent with Halley's measurements within uncertainty and still at least factor of two larger than for 67P dust. This suggests that kg-sized cometary meteoroids either contain more hydrogen than spacecraft-sampled dust particles, or actual plasma conditions differ significantly from our defaults (T = $10500$ K, $n_e = 4.8 \times 10^{13}$ cm$^{-3}$). Higher H/Si in larger meteoroids may reflect origin from comet interiors less affected by space weathering and volatile loss compared to surface dust. The mass-dependent variation supports our interpretation that smaller meteoroids experience more efficient hydrogen depletion during interplanetary transport, while larger bodies better retain volatiles, making cometary material an important contributor to Earth's volatile inventory.

The COSIMA instrument results from 67P indicate that cometary organic matter contains higher hydrogen content than meteoritic counterparts \citep{fray2016nitrogen}, consistent with our observation of lower H/Si in asteroidal meteors compared to cometary material of equivalent mass.

\subsubsection{Meteorite composition}

Hydrogen abundances in meteorites reflect volatile depletion patterns from early solar system formation. CI carbonaceous chondrites represent the most primitive and hydrogen-rich materials available for study, with up to 20 wt\% H$_2$O \citep{palme2022composition}. Multiple literature compilations of CI chondrite H/Si values by \cite{Wasson1988}, \cite{AndersGrevesse1989}, \cite{Lodders2009}, and \cite{Lodders2020} converge on values that closely match the average for our asteroidal meteor sample, despite the small number of such meteors with detectable hydrogen that is reflected in relatively high uncertainty of the average value (see Figure~\ref{FigH_Si_Comparison} and in Table~\ref{tab:hsi_comparison}).

\cite{Wasson1988} also published abundance values for other carbonaceous chondrites: CM, CO, CV. When converted from mass abundances to atomic abundances we receive lower values of H/Si as it can be seen in Figure \ref{FigH_Si_Comparison} and Table~\ref{tab:hsi_comparison}. Nevertheless, our observations suggest that detection of H$_\alpha$ line in the spectrum of meteor with asteroidal origin can indicate carbonaceous chondrite.

\cite{alexander2013classification} established that hydrogen abundances relative to refractory elements serve as indicators of aqueous alteration in carbonaceous chondrites, with H/Si values increasing systematically during phyllosilicate formation. Given the typical difficulty of detecting hydrogen in low-velocity asteroidal meteor spectra, the presence of H$_\alpha$ in our asteroidal sample may indicate enhanced hydrogen retention from aqueous alteration processes in their parent bodies, making H/Si values a powerful diagnostic tool for identifying the degree of parent body processing in observed meteoroids.

\subsection{Implications for Solar System water delivery models}

Our findings of variable hydrogen content in cometary meteoroids, particularly the correlation between H/Si and meteoroid mass, have implications for models of volatile delivery to the early Earth and inner Solar System.

The origin of Earth's water remains a fundamental question in planetary science. Current models generally fall into three categories: (1) water was delivered by comets, (2) water was delivered by water-rich carbonaceous asteroids, or (3) Earth accreted with its water already incorporated. Our results support models where comets could be significant contributors of volatiles to Earth. While isotopic measurements (particularly D/H) have challenged the role of comets in Earth's water delivery \citep{Altwegg2015}, the more recent finding of lower D/H outside cometary dust \citep{Mandt2024} reopens this possibility. Our observation of preserved high H/Si in larger meteoroids suggests that primitive, volatile-rich cometary material can survive at $1$~AU and could have delivered hydrogen and other volatiles to early Earth.

However, our results suggest that the volatile content of cometary material is complex and variable. The higher H/Si observed in larger meteoroids indicate that smaller cometary fragments might not fully represent the bulk composition of their parent bodies. This size-dependent effect could have consequences for interpreting the composition of interplanetary dust particles of cometary origin.

\section{Conclusions}\label{conclusions}

Our spectroscopic study of high-temperature components in meteor fireballs has revealed several key findings regarding the abundance and distribution of hydrogen and silicon in meteoroids of different origins:

1. The H-to-Si ratio shows no correlation with meteor velocity, but exhibits a clear positive correlation with photometric mass for cometary meteoroids. The correlation between meteoroid mass and H/Si  persists across different meteor showers, suggesting a fundamental physical process related to volatile preservation that depends on body size rather than on specific parent body composition. We suggest that larger bodies better preserve their volatile content.

2. Video spectroscopy confirms that afterglow contamination contributes to H$_\alpha$ and Si~II~-~2 lines, but does not fully explain the elevated hydrogen signals observed in some meteors. The very bright Lyrid meteor maintained a high H/Si value even after correction for afterglow contamination.

3. Our observations suggest that the abundance of hydrogen in cometary meteoroids is higher than in CI chondrites. Moreover, in meteoroids with masses exceeding $\approx$~0.5~kg it can be higher than the measured abundance in dust particles of comet Halley when temperature $10500$~K and electron density $4.8 \times 10^{13}$ cm$^{-3}$ is considered for the high-temperature component of meteor plasma. Comparable H/Si abundances were achieved when slightly a lower temperature or one-half the electron density was assumed. Nevertheless, estimated H/Si values in cometary meteoroids support models in which comets could be significant contributors to Earth's volatile inventory, particularly in light of recent findings showing more Earth-like D/H values in cometary gas compared to dust.

Future studies would benefit from observations of larger meteoroids across a wider range of meteor showers, spectroscopic techniques that can better separate afterglow from meteor head emissions, high-resolution spectroscopy, and coordinated measurements that combine spectroscopic data with precise orbital and physical parameter determinations.

\begin{acknowledgements}

This work was supported by the grant 24-10143S of the Grant Agency of the Czech Republic, GA \v CR.

\end{acknowledgements}

\bibliographystyle{aa} 
\bibliography{mojedatabase}

\begin{appendix}

\onecolumn
\section{Properties of studied meteor sample}
\nopagebreak[4]

\begin{table*}[htbp]
\centering
\caption{Meteors studied in this work and their properties. }
\label{tab:data}
\setlength{\tabcolsep}{4pt}
\small 
\begin{tabular}{l c r r r r r r r r }
\hline
ID               & Shower       & Mag    &   v        &  PhMass           & H I - 1/Si II - 2 & H I/Si II         & H/Si                & Perihelion  & Tj\\
                  &                   &             & (km/s)  &  (kg)                  &                   &                   &                     &        (AU) &      \\
\hline                                                                                                               
EN311216\_211702  & Quadrant.    & -11.51 & 41.50      & 3.0  $\times$10$^{-1}$ &           0.33        &         0.51      &          6.69       &       0.97  & 1.98   \\
EN270217\_023122  & sporadic     & -12.38 & 30.85      & 3.9  $\times$10$^{0} $ &   $\le$   0.07        &  $\le$  0.11      &  $\le$   1.39       &       0.36  & 3.02   \\
EN151117\_022348  & Leonids      & -12.01 & 71.56      & 1.3  $\times$10$^{-2}$ &           0.38        &         0.58      &          7.69       &       0.98  & -0.63  \\
EN241117\_024100  & sporadic     & -12.38 & 70.46      & 2.1  $\times$10$^{-2}$ &   $\le$   0.20        &  $\le$  0.30      &  $\le$   3.98       &       0.92  & -0.70  \\
EN050718\_231248  & sporadic     & -13.52 & 49.69      & 2.2  $\times$10$^{-1}$ &           0.84        &         1.28      &          16.78      &       0.57  & 0.24   \\
EN060818\_221424  & Perseids     & -11.45 & 59.86      & 2.6  $\times$10$^{-2}$ &   $\le$   0.11        &  $\le$  0.17      &  $\le$   2.17       &       0.96  & -0.44  \\
EN130818\_003733  & Perseids     & -10.28 & 59.92      & 3.1  $\times$10$^{-3}$ &   $\le$   0.20        &  $\le$  0.31      &  $\le$   4.04       &       0.95  & -0.04  \\
EN130818\_021607  & Perseids     & -10.00 & 61.06      & 2.0  $\times$10$^{-3}$ &   $\le$   0.23        &  $\le$  0.35      &  $\le$   4.56       &       0.95  & -0.22  \\
EN041118\_203933  & sporadic     & -13.60 & 29.22      & 1.4  $\times$10$^{0} $ &           0.97        &         1.48      &          19.43      &       0.86  & 1.16   \\
EN171118\_011528  & sporadic     & -13.22 & 69.52      & 1.8  $\times$10$^{-1}$ &           0.34        &         0.52      &          6.78       &       0.75  & -0.97  \\
EN181118\_040027  & Leonids      & -11.74 & 71.63      & 2.1  $\times$10$^{-2}$ &           0.09        &         0.13      &          1.76       &       0.98  & -0.67  \\
EN171119\_041459  & Leonids      & -15.96 & 71.18      & 1.5  $\times$10$^{0} $ &           1.21        &         1.84      &          24.15      &       0.98  & -0.73  \\
EN020820\_001158  & sporadic     & -14.30 & 66.47      & 2.3  $\times$10$^{-1}$ &           0.33        &         0.51      &          6.70       &       0.75  & -0.79  \\
EN140920\_025301  & sporadic     & -12.92 & 70.91      & 3.1  $\times$10$^{-2}$ &           0.25        &         0.37      &          4.92       &       0.87  & -0.97  \\
EN191120\_024524  & Leonids      & -10.12 & 71.87      & 4.5  $\times$10$^{-3}$ &   $\le$   0.09        &  $\le$  0.13      &  $\le$   1.74       &       0.99  & -0.73  \\
EN230421\_013038  & Lyrids       & -14.23 & 48.16      & 1.3  $\times$10$^{0} $ &           1.75        &         2.66      &          34.98      &       0.91  & 0.70   \\
EN060821\_222024  & sporadic     & -11.79 & 50.40      & 4.6  $\times$10$^{-2}$ &           0.27        &         0.41      &          5.43       &       0.89  & 0.46   \\
EN011021\_222435  & S. Taurids   & -10.64 & 31.18      & 1.1  $\times$10$^{-1}$ &   $\le$   0.33        &  $\le$  0.50      &  $\le$   6.63       &       0.31  & 3.32   \\
EN301022\_221704  & S. Taurids   & -12.71 & 32.93      & 9.1  $\times$10$^{-1}$ &   $\le$   0.65        &  $\le$  0.99      &  $\le$   12.96      &       0.29  & 2.96   \\
EN250123\_032350  & sporadic     & -12.52 & 28.48      & 1.1  $\times$10$^{1} $ &   $\le$   0.38        &  $\le$  0.58      &  $\le$   7.66       &       0.29  & 4.81   \\
EN090723\_232415  & sporadic     & -11.69 & 43.30      & 8.7  $\times$10$^{-2}$ &           0.61        &         0.93      &          12.25      &       1.02  & 0.72   \\
EN120923\_032745  & sporadic     & -12.71 & 67.84      & 5.3  $\times$10$^{-2}$ &   $\le$   0.17        &  $\le$  0.26      &  $\le$   3.39       &       0.60  & -0.93  \\
EN160923\_212646  & Sep. Eps. P. & -10.62 & 66.40      & 6.5  $\times$10$^{-3}$ &   $\le$   0.20        &  $\le$  0.30      &  $\le$   3.99       &       0.72  & -0.82  \\
EN100124\_183936  & sporadic     & -11.71 & 40.13      & 5.2  $\times$10$^{-2}$ &           0.40        &         0.61      &          8.06       &       0.25  & 0.65   \\
EN200324\_030519  & sporadic     & -12.97 & 68.98      & 2.4  $\times$10$^{-2}$ &           0.29        &         0.45      &          5.86       &       0.96  & -0.29  \\
EN290724\_223316  & sporadic     & -12.04 & 60.29      & 1.7  $\times$10$^{-2}$ &           0.57        &         0.87      &          11.49      &       0.97  & -0.43  \\
EN120824\_232436  & sporadic     & -10.56 & 55.62      & 3.9  $\times$10$^{-3}$ &           0.26        &         0.40      &          5.29       &       1.01  & -0.18  \\
EN130824\_000116  & Perseids     & -11.67 & 60.66      & 1.1  $\times$10$^{-2}$ &           0.51        &         0.78      &          10.20      &       0.97  & -0.28  \\
EN130824\_001932  & Perseids     & -10.19 & 60.73      & 2.1  $\times$10$^{-3}$ &   $\le$   0.10        &  $\le$  0.15      &  $\le$   1.95       &       0.95  & -0.33  \\
EN220924\_023610  & Sep. Eps. P. & -12.80 & 65.77      & 2.2  $\times$10$^{-2}$ &           0.53        &         0.81      &          10.63      &       0.67  & -0.76  \\
EN220425\_020108  & Lyrids       & -10.45 & 48.04      & 2.6  $\times$10$^{-2}$ &           0.06        &         0.09      &          1.15       &       0.92  & 0.39   \\
\hline
\end{tabular}
\tablefoot{The symbol $\le$ marks cases where $H_\alpha$ was not detected or with S/N close to $1$; $v$ is the velocity at the average trajectory point; $PhMass$ is the  photometric mass; the H I/Si II relative atomic abundance was computed for an assumed temperature of 10500~K; the $H/Si$ relative atomic abundance was computed for an assumed temperature of 10500~K and electron density $n_e = 4.8 \times 10^{13} \, \text{cm}^{-3}$.}

\end{table*}

\onecolumn
\section{Spectral region $600$~nm to $700$~nm of shower and asteroidal meteors}
\nopagebreak[4]

\begin{figure}[!htbp]
\centering
\begin{minipage}{.5\textwidth}
\centering
\includegraphics[width=.9\linewidth]{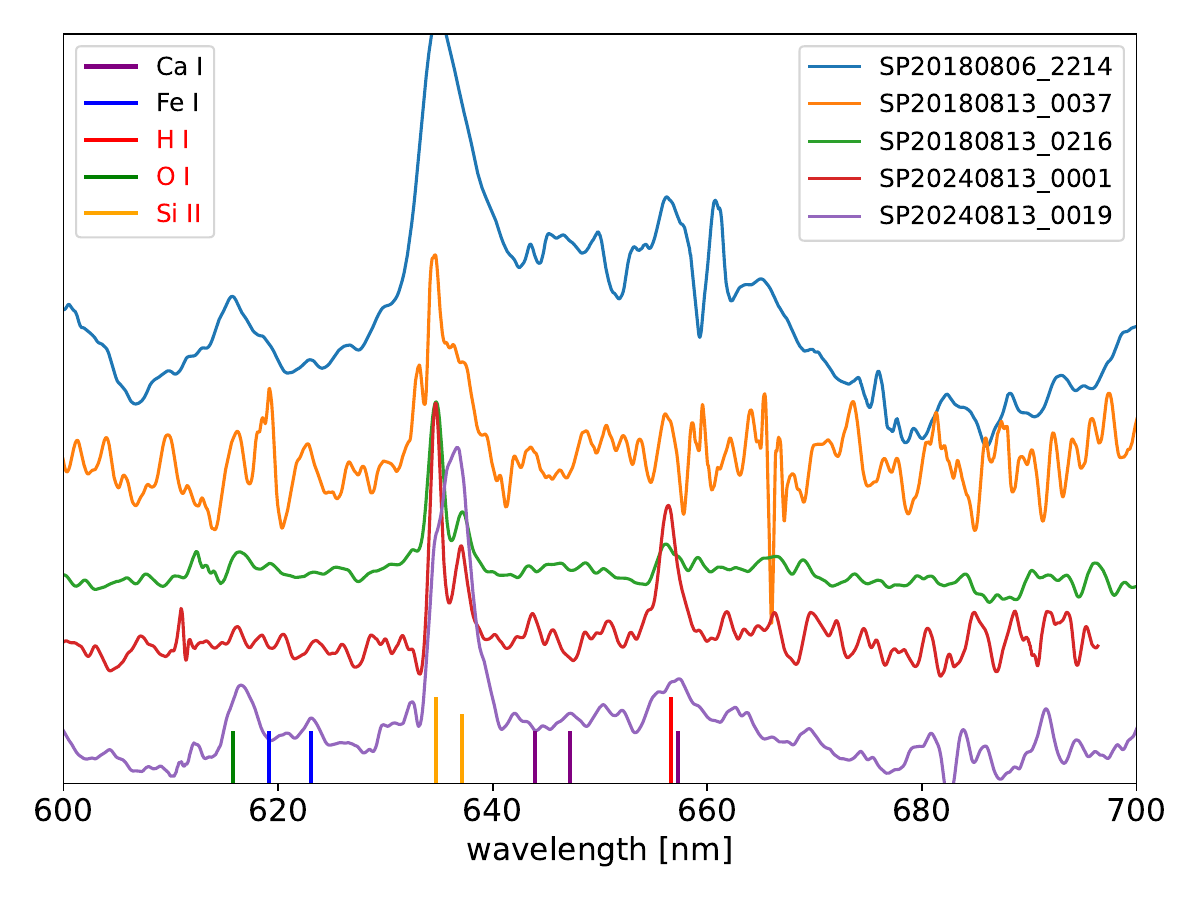}
\captionsetup{width=.85\linewidth}
\captionof{figure}{Spectral region of ionised silicon and neutral hydrogen for Perseid spectra. The theoretical positions of selected element lines are marked. The high-temperature component lines are shown in red font (see left inset).}
\label{FigPerSpectra}
\end{minipage}%
\begin{minipage}{.5\textwidth}
\centering
\includegraphics[width=.9\linewidth]{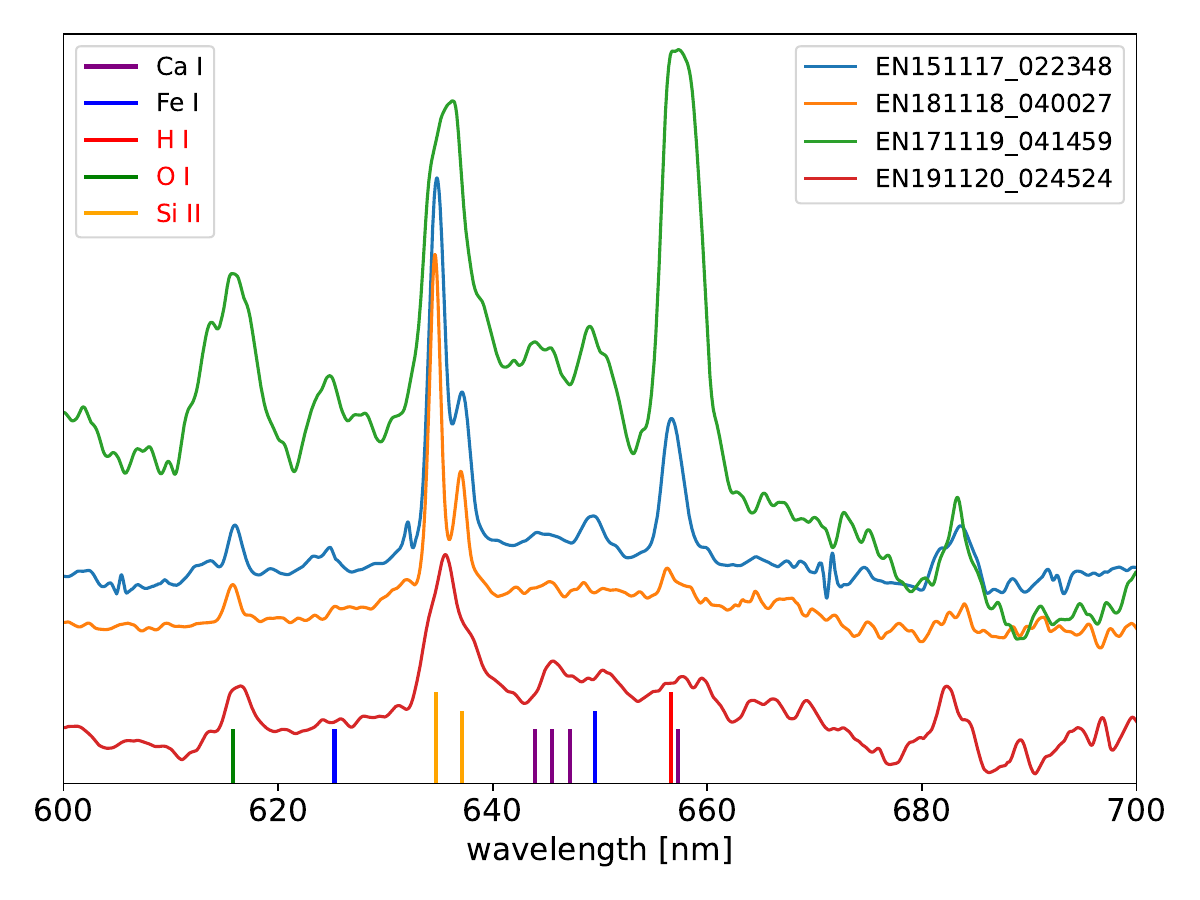}
\captionsetup{width=.85\linewidth}
\captionof{figure}{Spectral region of ionised silicon and neutral hydrogen for Leonid spectra. The theoretical positions of selected element lines are marked. The high-temperature component lines are shown in red font (see left inset).}
\label{FigLeoSpectra}
\end{minipage}

\begin{minipage}{.5\textwidth}
\centering
\includegraphics[width=.9\linewidth]{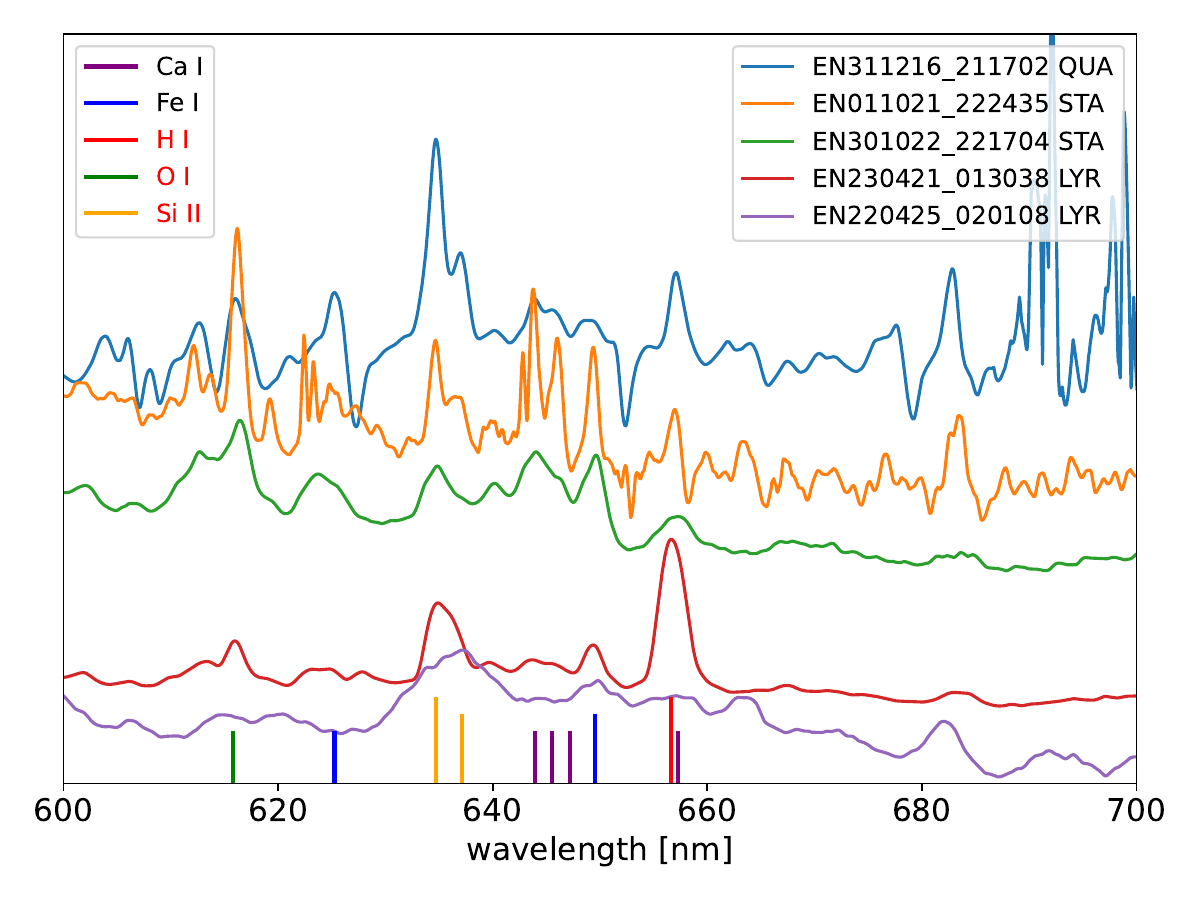}
\captionsetup{width=.85\linewidth}
\captionof{figure}{Spectral region of ionised silicon and neutral hydrogen for one Quadrantid, two Southern Taurid, and two Lyrid meteor spectra. The theoretical positions of selected element lines are marked. The high-temperature component lines are marked in red font (see left inset).}
\label{figQUA}
\end{minipage}%
\begin{minipage}{.5\textwidth}
\centering
\includegraphics[width=.9\linewidth]{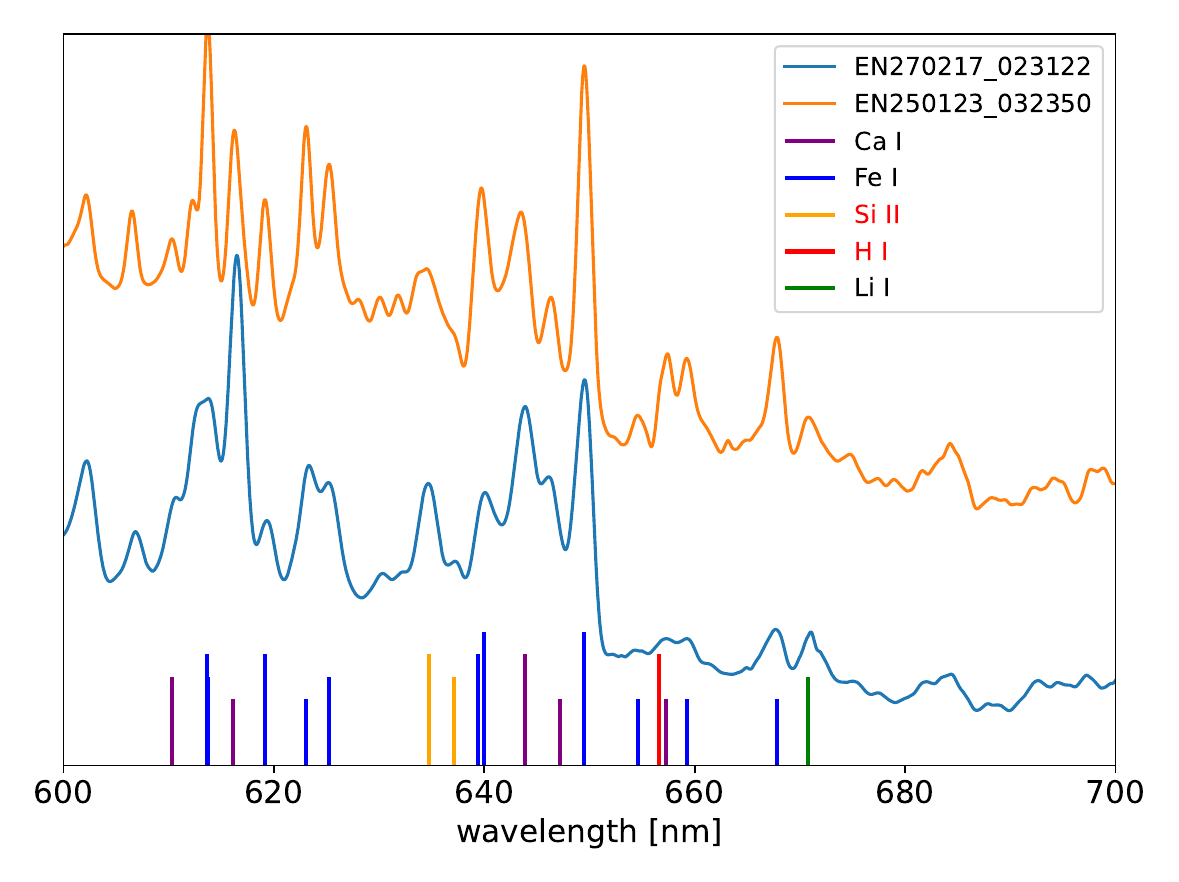}
\captionsetup{width=.85\linewidth}
\captionof{figure}{Spectral region of ionised silicon and neutral hydrogen for asteroidal-orbit meteor spectra. The theoretical positions of selected element lines are marked. The high-temperature component lines are marked in red font (see inset).}
\label{FigAST}
\end{minipage}
\end{figure}

\twocolumn
\section{Relative intensities of high-temperature component lines}
\nopagebreak[4]

\begin{figure}[!htbp]
\centering
\begin{subfigure}{\columnwidth}  
\includegraphics[width=\columnwidth]{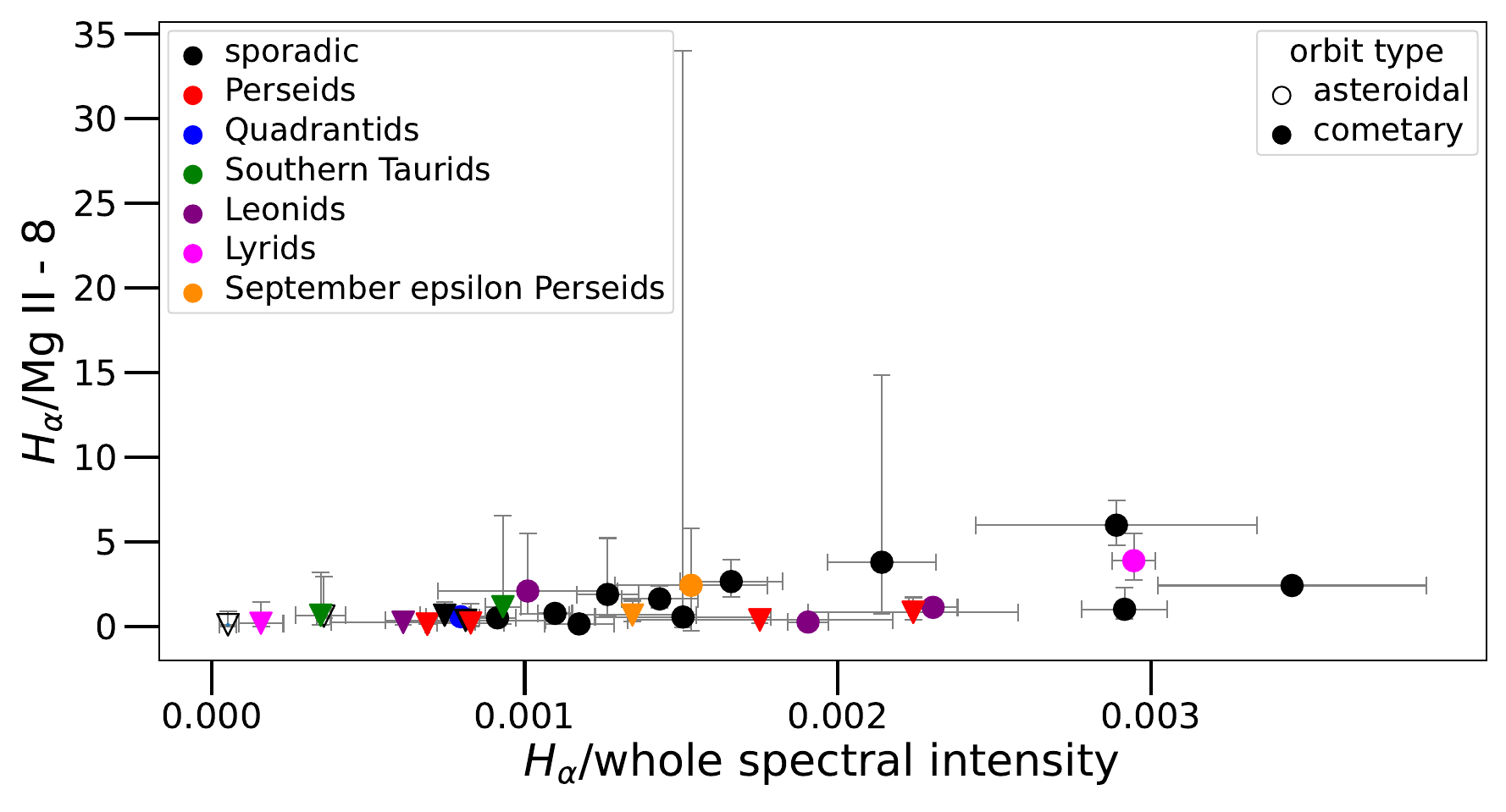}
\caption{}
\label{Fig:Mg28}
\end{subfigure}
\vfill
\begin{subfigure}{\columnwidth} 
\includegraphics[width=\columnwidth]{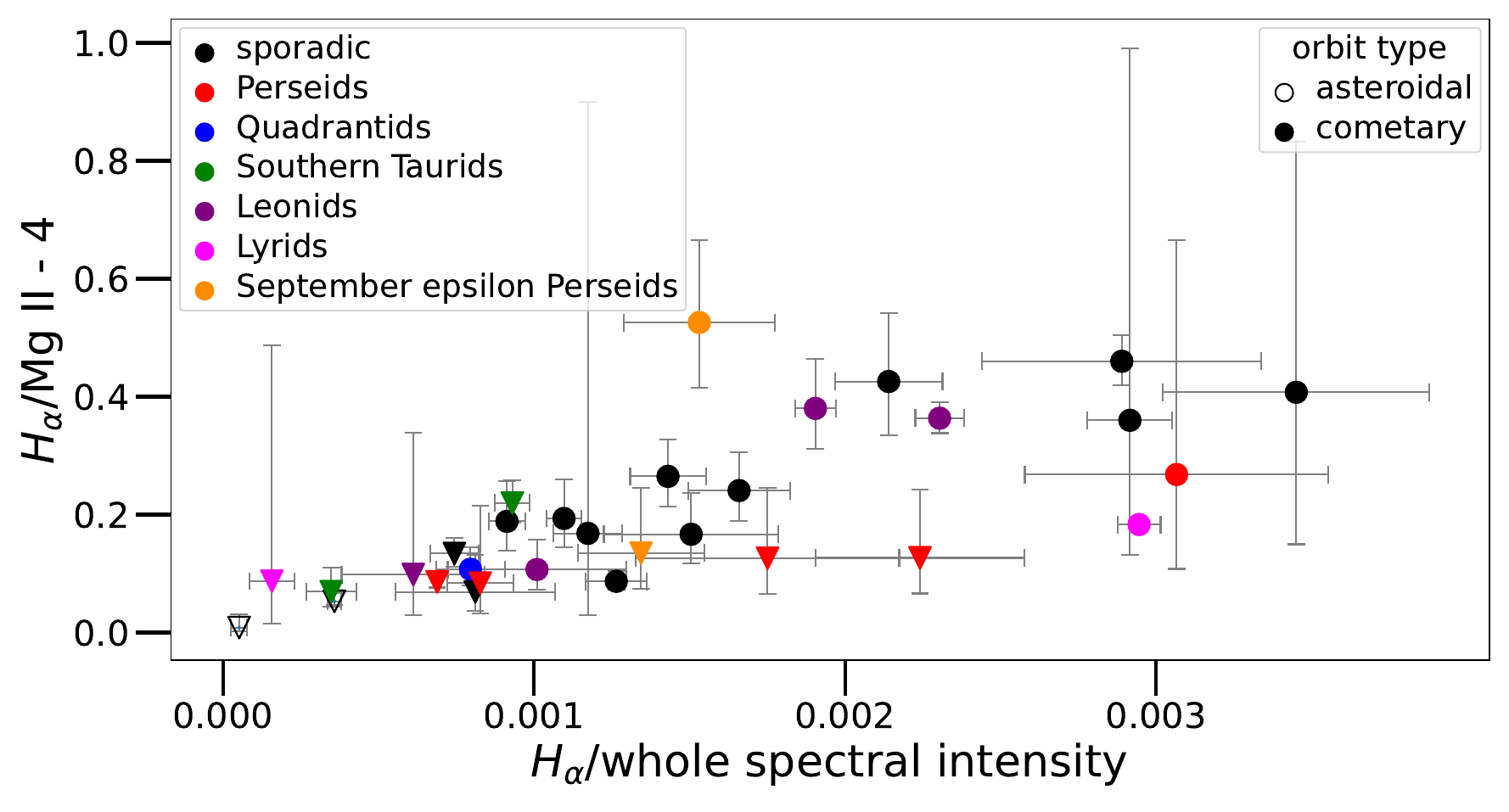}
\caption{}
\label{Fig:Mg24}
\end{subfigure}
\vfill
\begin{subfigure}{\columnwidth} 
\includegraphics[width=\columnwidth]{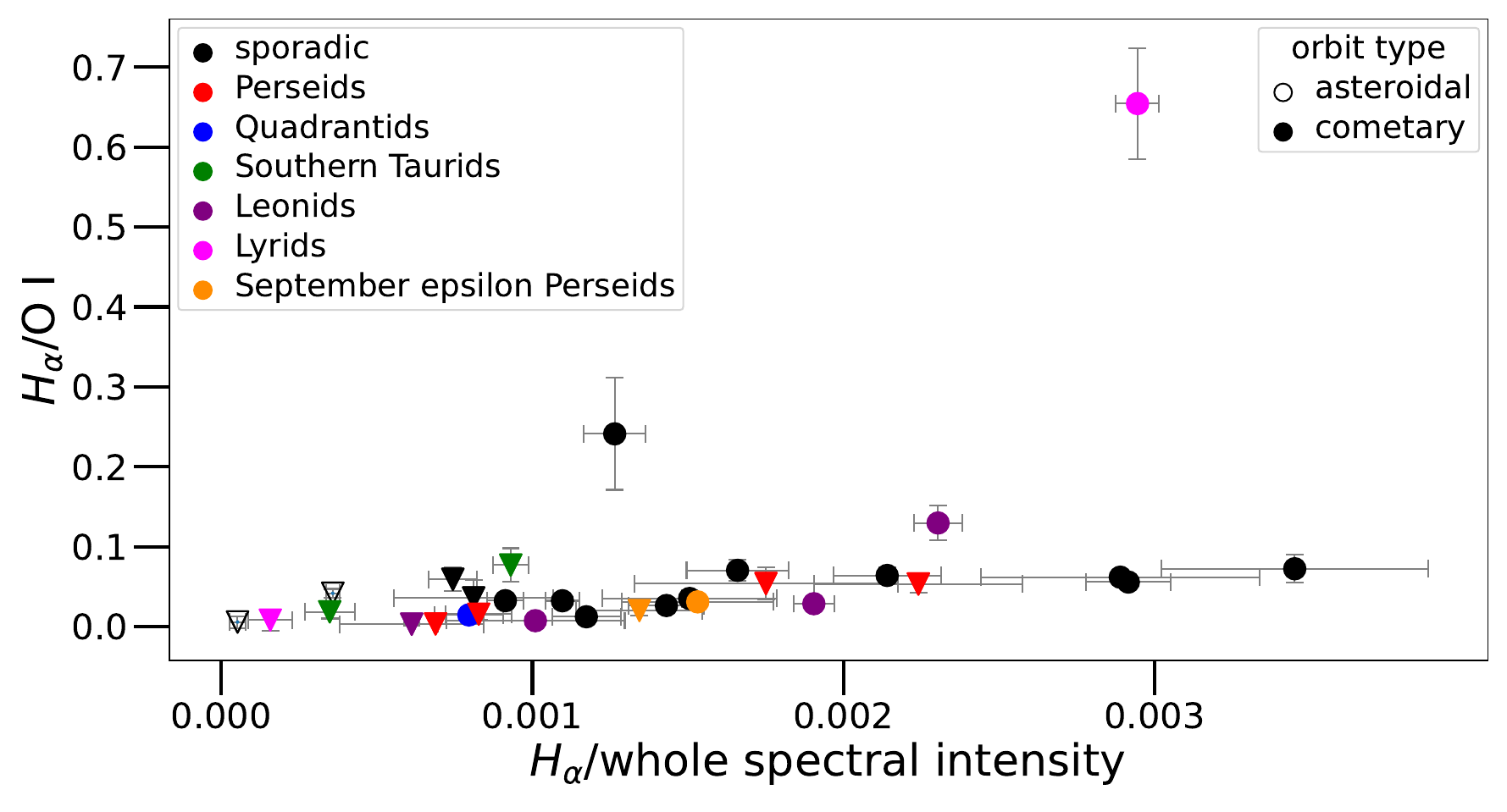}
\caption{}
\label{Fig:O}
\end{subfigure}
\caption{Relative intensities of high-temperature component lines. 
Meteors are marked as in Figure~\ref{FigHSiVel}. 
\textit{(a)} Ratio of the integrated intensity at H$_\alpha$ to Mg II–8 at 789.63 nm compared to the ratio of H$_\alpha$ to the whole observed spectrum. 
\textit{(b)} Ratio of the integrated intensity at H$_\alpha$ to Mg II–4 at 448 nm compared to the ratio of H$_\alpha$ to the whole observed spectrum. 
\textit{(c)} Ratio of the integrated intensity at H$_\alpha$ to O I–1 at 777 nm as a function of the ratio of the intensity in the H$_\alpha$ region to the intensity of the whole observed spectrum.}
\label{fig:MgOWholeEnergy}
\end{figure}

\section{Miscellaneous}
\nopagebreak[4]

\begin{figure}[!htbp]
\centering
\includegraphics[width=\columnwidth]{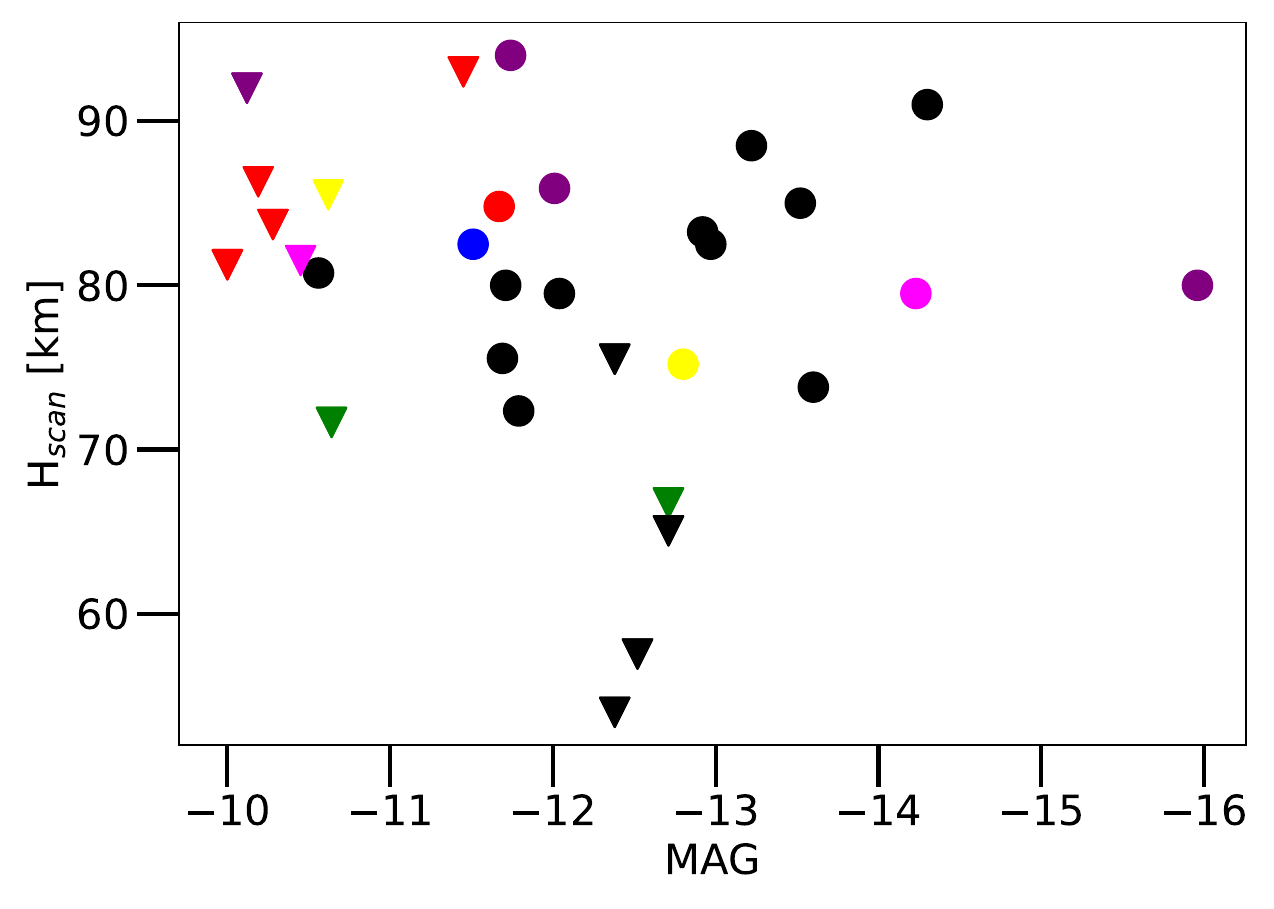}
\caption{Average altitude at which the measured spectrum was scanned vs meteor magnitude.}
\label{Fig:MagScan}
\end{figure}

\begin{table}[!htbp]
\centering
\caption{Percentage of neutral hydrogen atoms among all hydrogen atoms.}
\label{tab:hydrogen_compact}
\small
\resizebox{\columnwidth}{!}{%
\begin{tabular}{|c|cccc|}
\hline
\textbf{T [K]} & \multicolumn{4}{c|}{\textbf{Electron Density $n_e$ [cm$^{-3}$]}} \\
\hline
 & \textbf{2.5e13} & \textbf{5.0e13} & \textbf{7.5e13} & \textbf{1.0e14} \\
\hline
\textbf{10000} & 6.9 & 12.9 & 18.2 & 22.8 \\
\textbf{10100} & 5.9 & 11.1 & 15.8 & 20.0 \\
\textbf{10200} & 5.0 & 9.5 & 13.7 & 17.4 \\
\textbf{10300} & 4.3 & 8.2 & 11.8 & 15.2 \\
\textbf{10400} & 3.7 & 7.1 & 10.2 & 13.2 \\
\textbf{10500} & 3.1 & 6.1 & 8.9 & 11.5 \\
\textbf{10600} & 2.7 & 5.3 & 7.7 & 10.0 \\
\textbf{10700} & 2.3 & 4.5 & 6.7 & 8.7 \\
\textbf{10800} & 2.0 & 3.9 & 5.8 & 7.6 \\
\textbf{10900} & 1.7 & 3.4 & 5.0 & 6.6 \\
\textbf{11000} & 1.5 & 3.0 & 4.4 & 5.8 \\
\hline
\end{tabular}%
}
\end{table}

\begin{table}[!htbp]
\centering
\caption{Percentage of singly ionised silicon atoms among all silicon atoms.}
\label{tab:silicon_compact}
\small
\resizebox{\columnwidth}{!}{%
\begin{tabular}{|c|cccc|}
\hline
\textbf{T [K]} & \multicolumn{4}{c|}{\textbf{Electron Density $n_e$ [cm$^{-3}$]}} \\
\hline
 & \textbf{2.5e13} & \textbf{5.0e13} & \textbf{7.5e13} & \textbf{1.0e14} \\
\hline
\textbf{10000} & 84.0 & 91.3 & 94.0 & 95.4 \\
\textbf{10100} & 81.1 & 89.5 & 92.8 & 94.5 \\
\textbf{10200} & 77.9 & 87.5 & 91.3 & 93.3 \\
\textbf{10300} & 74.3 & 85.3 & 89.7 & 92.0 \\
\textbf{10400} & 70.5 & 82.7 & 87.7 & 90.5 \\
\textbf{10500} & 66.5 & 79.8 & 85.6 & 88.8 \\
\textbf{10600} & 62.2 & 76.7 & 83.2 & 86.8 \\
\textbf{10700} & 57.9 & 73.3 & 80.5 & 84.6 \\
\textbf{10800} & 53.5 & 69.7 & 77.5 & 82.1 \\
\textbf{10900} & 49.1 & 65.9 & 74.3 & 79.4 \\
\textbf{11000} & 44.8 & 61.9 & 70.9 & 76.5 \\
\hline
\end{tabular}%
}
\end{table}

\onecolumn
\section{Comparison with literature}
\nopagebreak[4]

\begin{table}[htbp]
\centering
\caption{H/Si atomic number ratio comparison between this work and the literature.}
\label{tab:hsi_comparison}
\footnotesize
\renewcommand{\arraystretch}{1.3}
\begin{tabular}{@{}p{3.5cm}p{1.6cm}p{1.6cm}p{1.4cm}p{1.4cm}p{1.4cm}p{1.4cm}p{0.6cm}p{1.8cm}@{}}
\toprule
\multicolumn{9}{c}{\textbf{This study}} \\
\midrule
{\textbf{Mass Range}} & \multicolumn{6}{c}{\textbf{H/Si}} & \multirow{2}{*}{\textbf{N}}  & \multirow{2}{*}{\textbf{Notes}} \\
\cmidrule(lr){2-7}
\shortstack{T (K)\\$n_e (cm^{-3})$} 
& \textbf{Combined} 
& \multicolumn{1}{r}{\shortstack{10500\\$4.8 \times 10^{13}$}} 
& \shortstack{10000\\$4.8 \times 10^{13}$} 
& \shortstack{11000\\$4.8 \times 10^{13}$} 
& \shortstack{10500\\$1.0 \times 10^{14}$} 
& \shortstack{10500\\$2.5 \times 10^{13}$} 
&  & \\
\midrule

\shortstack{Smallest meteoroids \\($<$0.01~kg)} 
& \textbf{$3.6 \pm 1.5$} 
& $3.6_{-0.7}^{+0.9}$ & $2.2_{-0.4}^{+0.5}$ & $5.1_{-1.0}^{+1.2}$ & $2.1_{-0.4}^{+0.5}$ & $5.6_{-1.1}^{+1.3}$ & 6 & cometary \\

\shortstack{Small meteoroids \\(0.01--0.1~kg)}  
& \textbf{$6.4 \pm 3.0$} 
& $6.4_{-1.0}^{+1.0}$ & $3.9_{-0.6}^{+0.6}$ & $9.1_{-1.4}^{+1.4}$ & $3.6_{-0.6}^{+0.6}$ & $10.0_{-1.6}^{+1.6}$ & 14 & cometary \\

\shortstack{Medium meteoroids \\(0.1--0.5~kg)}  
& \textbf{$8.7 \pm 4.0$} 
& $8.7_{-2.1}^{+2.2}$ & $5.3_{-1.3}^{+1.3}$ & $12.5_{-3.0}^{+3.1}$ & $5.0_{-1.2}^{+1.3}$ & $13.7_{-3.3}^{+3.4}$ & 5 & cometary \\

\shortstack{Large meteoroids \\($>$0.5~kg)}  
& \textbf{$23 \pm 10$} 
& $23_{-5}^{+5}$ & $13.9_{-3.0}^{+3.1}$ & $33_{-7}^{+7}$ & $13.1_{-2.8}^{+2.9}$ & $36_{-8}^{+8}$ & 6 & cometary \\

\shortstack{Asteroidal \\   \smallskip} 
& \textbf{$4.5 \pm 3.5$} 
& $4.5_{-3.2}^{+4.0}$ & $2.7_{-2.0}^{+2.4}$ & $6.5_{-4.6}^{+5.7}$ & $2.6_{-1.8}^{+2.3}$ & $7.1_{-5.0}^{+6.2}$ & 2 & asteroidal \\

\midrule
\multicolumn{9}{c}{\textbf{Literature}} \\
\midrule

\multicolumn{2}{l}{\textbf{Category }} & \textbf{H/Si} &  \multicolumn{1}{c}{\textbf{N}} & \multicolumn{2}{l}{\textbf{Source/Reference}} & \multicolumn{3}{l}{\textbf{Notes}} \\
\midrule

\multicolumn{2}{l}{Meteors - Perseids} & 0.8 \& 1.5 &  \multicolumn{1}{c}{2} & \multicolumn{2}{l}{Borovička \& Betlem, 1997} & \multicolumn{3}{l}{photographic spectra, H$_\beta$} \\
\multicolumn{2}{l}{Meteors - Perseids and Leonids} & 1.5--21.4 &  \multicolumn{1}{c}{9} & \multicolumn{2}{l}{Borovička et al., 2004} & \multicolumn{3}{l}{photographic spectra} \\
\multicolumn{2}{l}{Meteors - Leonid} & 4.4 &  \multicolumn{1}{c}{1} & \multicolumn{2}{l}{Jenniskens \& Mandell, 2004} & \multicolumn{3}{l}{video spectrum} \\
\midrule

\multicolumn{2}{l}{Comets - Halley's dust} & $10.95 \pm 2.37$ &  \multicolumn{1}{c}{--} & \multicolumn{2}{l}{Jessberger, 1988} & \multicolumn{3}{l}{PUMA--1 mass spectrometer} \\
\multicolumn{2}{l}{\shortstack{Comets - 67P/Churyumov\\--Gerasimenko dust}} & $5.72_{-1.25}^{+1.46}$ & \multicolumn{1}{c}{--} & \multicolumn{2}{l}{Engrand et al., 2024} & \multicolumn{3}{l}{COSIMA instrument} \\
\midrule

\multicolumn{2}{l}{Meteorites - CI chondrites} & 5.308 &\multicolumn{1}{c}{--} & \multicolumn{2}{l}{Wasson \& Kallemeyn, 1988} & \multicolumn{3}{l}{} \\
\multicolumn{2}{l}{Meteorites - CI chondrites} & 5.29 & \multicolumn{1}{c}{--} & \multicolumn{2}{l}{Anders \& Grevesse, 1989} & \multicolumn{3}{l}{compilation} \\
\multicolumn{2}{l}{Meteorites - CI chondrites} & 5.13 & \multicolumn{1}{c}{--} & \multicolumn{2}{l}{Lodders et al., 2009} & \multicolumn{3}{l}{} \\
\multicolumn{2}{l}{Meteorites - CI chondrites} & $4.81 \pm 0.94$ & \multicolumn{1}{c}{--} & \multicolumn{2}{l}{Lodders, 2020} & \multicolumn{3}{l}{} \\
\multicolumn{2}{l}{Meteorites - CM chondrites} & 3.024 & \multicolumn{1}{c}{--} & \multicolumn{2}{l}{Wasson \& Kallemeyn, 1988} & \multicolumn{3}{l}{} \\
\multicolumn{2}{l}{Meteorites - CV chondrites} & 0.50 & \multicolumn{1}{c}{--} & \multicolumn{2}{l}{Wasson \& Kallemeyn, 1988} & \multicolumn{3}{l}{} \\
\multicolumn{2}{l}{Meteorites - CO chondrites} & 0.12 & \multicolumn{1}{c}{--} & \multicolumn{2}{l}{Wasson \& Kallemeyn, 1988} & \multicolumn{3}{l}{} \\

\bottomrule
\end{tabular}
\tablefoot{ The combined values correspond to the adopted plasma parameters (10500 K, $4.8 \times 10^{13}$ $\mathrm{cm}^{-3}$). Their quoted uncertainties combine the measurement and statistical errors of the adopted case with the variation across all tested plasma conditions. The uncertainties of the other values from this study are a combination of the measurement and statistical errors.}
\end{table}

\end{appendix}

\end{document}